\documentclass[amsmath,amssymb,prb]{revtex4}

\usepackage{graphics}

\begin{document}
\title{Many-body theory of degenerate systems}
\author{Christian Brouder}
\affiliation{%
Laboratoire de Min\'eralogie Cristallographie, CNRS UMR 7590,
Universit\'es Paris 6 et 7, IPGP, case 115, 4 place Jussieu,
75252 Paris cedex 05, France.
}%

\date{\today}

%
\begin{abstract}
The many-body theory of degenerate systems is described in detail.
More generally, this theory applies to
systems with an initial state that cannot be described by 
a single Slater determinant. The double-source (or closed-time-path)
formalism of nonequilibrium quantum field theory is
used to derive an expression for the average value
of a product of interacting fields when the initial
state is not the vacuum or a Slater determinant.
Quantum group techniques are applied to derive
the hierarchy of unconnected Green functions and the hierarchy of
connected ones. 
\end{abstract}
%
\maketitle
%

\newcommand{\un}{{\mathbf{1}}}

\newcommand{\ee}{{\mathrm{e}}}
\newcommand{\tr}{{\mathrm{tr}}}
\newcommand{\dd}{{\mathrm{d}}}
\newcommand{\Id}{{\mathrm{Id}}}
\newcommand{\inter}{{\mathrm{int}}}
\newcommand{\degres}{{\mathrm{deg}}}
\newcommand{\barT}{{T^*}}
\newcommand{\calA}{{\cal{A}}}
\newcommand{\calD}{{\cal{D}}}
\newcommand{\calL}{{\cal{L}}_I}
\newcommand{\calH}{{\cal{H}}}
\newcommand{\action}{{\cal{A}^\inter}}
\newcommand{\counit}{{\varepsilon}}
\newcommand{\deltaj}{{\delta j}}
\newcommand{\barun}{{\bar{1}}}
\newcommand{\barunp}{{\bar{1}'}}
\newcommand{\barn}{{\bar{n}}}
\newcommand{\barnp}{{\bar{n}'}}
\newcommand{\barde}{{\bar{2}}}
\newcommand{\bardep}{{\bar{2}'}}
\newcommand{\barx}{{\bar{x}}}
\newcommand{\barxp}{{\bar{x}'}}
\newcommand{\baru}{{\bar{u}}}
\newcommand{\barv}{{\bar{v}}}
\newcommand{\bara}{{\bar{\alpha}}}
\newcommand{\bareta}{{\bar{\eta}}}
\newcommand{\barpsi}{{\bar{\psi}}}

\newcommand{\bfn}{{\mathbf{n}}}
\newcommand{\bfk}{{\mathbf{k}}}
\newcommand{\bfr}{{\mathbf{r}}}
\newcommand{\bfun}{{\mathbf{1}}}
\newcommand{\bfX}{{\mathbf{X}}}
\newcommand{\bflambda}{{\boldsymbol{\lambda}}}
\newcommand{\bfmu}{{\boldsymbol{\mu}}}

\renewcommand{\i}[1]{{}_{\scriptscriptstyle(#1)}}
\newcommand{\iu}[1]{{}_{\scriptscriptstyle(\underline #1)}}
\newcommand{\Deltau}{\underline{\Delta}}

\newcommand{\bc}{{\bar{c}}}
\newcommand{\bd}{{\bar{d}}}

\section{Introduction}
Degenerate systems are plenty
(all systems containing an odd number of
electrons, by Kramers theorem \cite{Kramers30,Sachs}) and 
degeneracy plays an important role in
numerous interesting physical effects
(e.g. magnetism or superconductivity).
Thus, it seems relevant to develop
calculation methods for degenerate
systems. 
The density functional theory of degenerate
systems is a subject of continued interest
\cite{LevyNagy,Yang,UllrichKohn1,UllrichKohn2,Sahni}.
Green functions can be useful because they provide
an exact expression for the exchange and correlation
potential \cite{Sham,Godby} and for their
ability to calculate excitations, e.g. through the 
GW approximation \cite{Arya} or
Bethe-Salpeter equation \cite{Rohlfing2000,Onida}.
Here, we study the Green functions of degenerate systems.

The quantum field theory of degenerate systems 
has been investigated since the sixties
by two methods: either by calculating 
the $S$-matrix elements between different
``in" and ``out" states 
\cite{Morita,Brandow,Oberlechner,Johnson,Dahmen}
or by assuming that the interacting
ground state is a pure state evolving from
a non-interacting pure state
\cite{BH58,Esterling,Kummel,Kuo,LindgrenMorrison}. 
However, these works treat the electron-electron
interaction as a perturbation, and we know
that this is a rather crude approximation.
So we need a non-perturbative approach
to the Green functions of degenerate systems.
The simplest and most common non-perturbative
method is self-consistency.
Therefore, we shall develop in this paper
a self-consistent calculation the Green
functions of a degenerate system.

The first problem that we meet with such a
program is the fact that we cannot describe
the system with a wavefunction. 
For example, the electronic configuration of a boron atom
in the ground state is 1s$^2$2s$^2$2p$^1$.
The Hamiltonian is invariant by rotation and,
if the spin-orbit coupling is neglected, 
the six pure states $|p_i,s\rangle$
(with $i=x,y,z$ and $s=\pm1/2$)
are degenerate and are eigenstates
of $L^2$ and $s^2$.
However, none of these pure states 
gives a spherically symmetric electron density.
More generally, if the ground state of a quantum
system is a pure state with angular momentum $L\ge1$,
the charge density derived from this state
is not spherically symmetric \cite{Lieb}.
Therefore, the self-consistent pure state
(with $L\ge1$) of a spherical Hamiltonian breaks
the spherical symmetry of the problem.
A related results was proved by Bach et al. \cite{Bach}:
the solution of the unrestricted Hartree-Fock
equations does not contain unfilled shells.
Therefore, for the boron atom, the
$2p$ shell is deformed to lift 
its degeneracy.

To cure this defect, we have to assume that
the boron atom is in the mixed state described
by the density matrix
$\hat\rho=(1/6)\sum_{i,s} |p_i,s\rangle\langle p_i,s|$,
which preserves
the rotational symmetry of the system.

Therefore, for degenerate systems, we need to calculate
an evolution starting not from a single pure state
(usually called the vacuum $|0\rangle$),
but from a density matrix $\hat\rho$.
The best tool to do so is nonequilibrium quantum
field theory, as created by
Schwinger, Kadanoff, Baym and Keldysh 
\cite{SchwingerJMP,Kadanoff,Keldysh}.
In particular, the closed time-path method
will enable us to express the various
Green functions we need as functional
derivatives with respect to external sources.

We describe now the main result of the paper.
The differential form of the
Green function hierarchy
\cite{MartinSchwinger,Kadanoff} is
\begin{eqnarray*}
\big( i \frac{\partial}{\partial t_1}
+ \frac{\Delta_{1}}{2m}\big)
G(1,1') &=&
\delta(1-1') - i
\int v(\bfr_1-\bfr_{2})G_2(1,2;1',2^+) \dd \bfr_2.
\end{eqnarray*}
For non-degenerate systems that can be represented
by a Slater determinant, the integral form of this equation is
\begin{eqnarray*}
G(1,1') &=&
G_0(1,1') - i
\int 
G_0(1,3)v(\bfr_3-\bfr_{2})G_2(3,2;1',2^+) 
\dd \bfr_2\dd \bfr_3,
\end{eqnarray*}
where $G_0(1,1')$ is the Green function
for the free Schr\"odinger equation
\begin{eqnarray*}
\big( i \frac{\partial}{\partial t_1}
+ \frac{\Delta_{1}}{2m}\big)
G_0(1,1') &=&
\delta(1-1').
\end{eqnarray*}
The unperturbed Green function $G^0(x,y)$ is given by
the following expression (see \cite{Fetter} p.124)
\begin{eqnarray}
G_0(x,x') &=& -i\theta(t-t')\sum_{\epsilon_n>\epsilon_F} 
\ee^{-i\epsilon_n(t-t')} u_n(\bfr) \baru_n(\bfr')
\nonumber\\&&
+i\theta(t'-t)\sum_{\epsilon_n\le\epsilon_F} \ee^{-i\epsilon_n(t-t')}
u_n(\bfr) \baru_n(\bfr'),
\label{defG0}
\end{eqnarray}
with $x=(t,\bfr)$, $x'=(t',\bfr')$ and the orbital
$u_n(\bfr)$ is the solution of the unperturbed
Schr\"odinger equation for energy $\epsilon_n$.
The Fermi energy $\epsilon_F$ is chosen so that the
total charge $-i\int\tr(G_0(t,\bfr,t,\bfr))\dd\bfr$ is equal
to the number $N$ of electrons in the system.
In this independent-particle picture, the ground
state is degenerate if the Fermi level is degenerate
and not completely filled. Therefore, the definition
(\ref{defG0}) must be modified because it assumes
that the Fermi level is full. We shall see that
the definition of $G_0$ for a degenerate system
is non trivial.

Moreover, for degenerate systems,
the relation between the differential
and the integral Kadanoff-Baym equations
is modified, because the solutions of the free Schr\"odinger
equation intervene. In fact, the
full hierarchy of Green functions is changed.

The correct hierarchy of Green function is
important because it is the basis of
the GW approximation \cite{Arya}.
Thus, the GW approximation must be adapted
to degenerate systems. A similar modification
is required for the Bethe-Salpeter equation.

In this paper, we give the proper expression
 for $G_0$ and the integral
form of the Kadanoff-Baym equation 
for a general density matrix.
Compared to previous results, the present
equations have two advantages: they 
are adapted to a self-consistent treatment
and they do not break the symmetry
of the problem.

Although the quantum field theory of degenerate
systems seems to be a rather natural problem,
it was not solved before because it poses
technical difficulties
that can hardly be overcome with the standard
many-body techniques. Our main tool here
will be the quantum group (or Hopf algebra) approach to 
quantum field theory, developed in
\cite{BrouderOecklI,BrouderQG}.

In this paper, we give a self-contained presentation
of the calculation of the expectation values of
products of quantum field operators in the
interaction representation. Then we compute
interacting Green functions using 
functional derivatives with respect
to external sources. The Hopf algebra of derivations
is then introduced and used to derive the
hierarchy of Green functions for systems
with degenerate initial states, or more
generally for systems with initial correlation.
Explicit hierarchies are obtained for unconnected
and connected Green functions.
In a forthcoming publication, the special case
of a single electron in a system with
closed shells and a two-fold degenerate orbital
will be calculated in detail.

\section{Evolution of expectations values}
We saw in the introduction that a self-consistent
calculation of degenerate systems requires the
use of density matrices.
In this section we investigate how
the unperturbed density matrix evolves
with time under perturbation.
As a first step, we calculate the evolution
of an unperturbed wavefunction, then
we extend this to the evolution of
a density matrix, and we use this
result to calculate the evolution
of an expectation value.
The calculation of transition amplitudes
in quantum field theory is not completely
standard, so we give here a detailed 
derivation.
As an application, we obtain a formula
for the Green function of a degenerate
system.

\subsection{Evolution of wavefunctions}
\label{evolutionwf}
\label{transampsect}

We start from a time-independent 
free Hamiltonian
$H_0=\int \barpsi_S(\bfr) h_0(\bfr) \psi_S(\bfr) \dd \bfr$,
where $h_0(\bfr)$ is a one-particle Hamiltonian
and $\psi_S(\bfr)$ is the field operator in the
Schr\"odinger picture.
A convient form of $\psi_S(\bfr)$ is
$\psi_S(\bfr) = \sum_n u_n(\bfr) b_n$,
where $u_n(\bfr)$ is an eigenstate of $h_0$:
$h_0 u_n=\epsilon_n u_n$, and $b_n$ is the
annihilation operator for the one-particle state $u_n(\bfr)$.
We first look for the solutions of the Schr\"odinger
equation
\begin{eqnarray*}
i\frac{\partial}{\partial t} |\Phi^0_n(t)\rangle
&=& H_0 |\Phi^0_n(t)\rangle.
\end{eqnarray*}
As usually, we isolate the time dependence by putting
$|\Phi^0_n(t)\rangle=\ee^{-iE^0_nt}|\Phi^0_n\rangle$
so that $H_0|\Phi^0_n\rangle=E^0_n|\Phi^0_n\rangle$.
We assume that the $|\Phi^0_n\rangle$ provide
a complete set of states.
The matrix elements of the operator
$A_S(t)$ in the Schr\"odinger picture 
is 
$\langle \Phi^0_m(t)| A_S(t)|\Phi^0_n(t)\rangle$.
We go now to the Heisenberg picture by
\begin{eqnarray*}
\langle \Phi^0_m(t)| A_S(t)|\Phi^0_n(t)\rangle
&=&
\langle \Phi^0_m|\ee^{iE^0_mt} A_S(t)\ee^{-iE^0_nt}|\Phi^0_n\rangle
\\&=&
\langle \Phi^0_m|\ee^{iH_0 t} A_S(t)\ee^{-i H_0 t}|\Phi^0_n\rangle
=
\langle \Phi^0_m|A(t)|\Phi^0_n\rangle,
\end{eqnarray*}
where
$A(t)=\ee^{iH_0 t} A_S(t)\ee^{-i H_0 t}$
is the operator $A_S(t)$ in the Heisenberg
picture.
In particular $\ee^{iH_0 t} H_0\ee^{-i H_0 t}=H_0$,
so that $H_0$ is the same in both picture.
The field operator in the Heisenberg picture
is (see \cite{Gross} p.146)
$\psi(x)=\ee^{iH_0 t} \psi_S(\bfr) \ee^{-i H_0 t}
=\sum_n u_n(\bfr) \ee^{-i \epsilon_n t} b_n$,
with $x=(t,\bfr)$.

We are interested in the interacting theory, so
we add a possibly time-dependent interaction term to the Hamiltonian.
This gives us $H_S(t)=H_0+H_S^\inter(t)$
in the Schr\"odinger picture and
$H(t)=H_0+H^\inter(t)$ in the Heisenberg picture,
with $H^\inter(t)=\ee^{iH_0 t} H^\inter_S(t) \ee^{-i H_0 t}$. 
In practice, $H^\inter_S(t)$ is a polynomial
in $\psi_S(\bfr)$ and $\barpsi_S(\bfr)$,
and $H^\inter(t)$ is the same polynomial
where $\psi_S(\bfr)$ is replaced by $\psi(t,\bfr)$
and $\barpsi_S(\bfr)$ is replaced by
$\barpsi(t,\bfr)$.

We look for solutions of the Schr\"odinger equation
\begin{eqnarray*}
i\frac{\partial}{\partial t} |\Phi^S_n(t)\rangle
&=& H_S(t) |\Phi^S_n(t)\rangle.
\end{eqnarray*}
We go to the Heisenberg reprensentation with
respect to $H_0$ (which is called the interaction
picture) and we define
$|\Phi_n(t)\rangle=\ee^{iH_0 t}|\Phi^S_n(t)\rangle$.
Therefore,
\begin{eqnarray*}
i\frac{\partial}{\partial t} |\Phi_n(t)\rangle
&=& \ee^{iH_0 t}(-H_0+H_S(t))|\Phi^S_n(t)\rangle
=H^\inter(t) |\Phi_n(t)\rangle.
\end{eqnarray*}
To solve this problem, we look for an operator
$V(t)$ such that $|\Phi_n(t)\rangle=V(t)|\Phi^0_n\rangle$.
The Schr\"odinger equation becomes
\begin{eqnarray*}
i\frac{\partial}{\partial t} V(t) |\Phi^0_n\rangle
&=& H^\inter(t) V(t)|\Phi^0_n\rangle.
\end{eqnarray*}
This must be true for the complete set of $|\Phi^0_n\rangle$,
thus 
\begin{eqnarray}
i\frac{\partial}{\partial t} V(t)
&=& H^\inter(t) V(t).
\label{dVdt}
\end{eqnarray}

\subsection{Calculation of $V(t)$}
To solve equation (\ref{dVdt}), we put
$U(t,t')=V(t) V^{-1}(t')$.
Therefore, $U(t,t)=1$ and
\begin{eqnarray*}
i\frac{\partial}{\partial t} U(t,t')
&=& H^\inter(t) U(t,t').
\end{eqnarray*}
We are going to prove some properties of 
$U(t,t')$. We first prove the group property
$U(t,t')U(t',t'')=U(t,t'')$.
From the fact that $V(t) V^{-1}(t)=1$
we deduce
\begin{eqnarray*}
i\frac{\partial}{\partial t} V^{-1}(t)
&=& -V^{-1}(t)H^\inter(t),
\end{eqnarray*}
and
\begin{eqnarray*}
i\frac{\partial}{\partial t'} U(t,t')
&=& -U(t,t')H^\inter(t'),
\end{eqnarray*}
Thus
\begin{eqnarray*}
i\frac{\partial}{\partial t'} U(t,t')U(t',t)
&=&
U(t,t')(-H^\inter(t')+H^\inter(t'))U(t',t)=0.
\end{eqnarray*}
Hence, the product $U(t,t')U(t',t)$
is independent of $t'$. To find its value,
we put $t'=t$, so that
$U(t,t')U(t',t'')=U(t,t)U(t,t'')=U(t,t'')$.
Then we show that $U(t,t')$ is unitary.
We take the adjoint of equation (\ref{dVdt}):
\begin{eqnarray*}
-i\frac{\partial}{\partial t} V^\dagger(t)
&=& V^\dagger(t)H^\inter(t),
\end{eqnarray*}
because $H^\inter(t)$ is Hermitian.
This implies
\begin{eqnarray*}
i\frac{\partial}{\partial t} U^\dagger(t,t')
&=&{V^{-1}}^\dagger(t')
i\frac{\partial}{\partial t} V^\dagger(t)
=
-U^\dagger(t,t') H^\inter(t).
\end{eqnarray*}
Therefore, $U^\dagger(t,t')=U(t',t)$ 
because both operators satisfy the same equation and
the same boundary condition $U(t,t)=U^\dagger(t,t)=1$.
But the group property leads to
$U(t,t')U(t',t)=U(t,t)=1$, so that
$U^\dagger(t,t')= U(t',t)=U^{-1}(t,t')$,
and $U(t,t')$ is unitary.

The construction of $U(t,t')$ is standard
(see, e.g. \cite{Dyson,Gross}) and yields
\begin{eqnarray}
U(t,t') &=& T \exp\big(-i\int_{t'}^t H^\inter(\tau) \dd \tau\big).
\label{U(tt')}
\end{eqnarray}
Here, $T$ is the time-ordering operator that orders its arguments
by decreasing time from left to right.
For example
$T(A(t)B(t'))$ is $A(t)B(t')$ if $t>t'$ and
is $B(t')A(t)$ if $t'>t$. An important property of the
time-ordering operator is that its arguments commute.
For instance, it can be checked from the definition
that $T(A(t)B(t'))=T(B(t')A(t))$.

To complete the picture, we use the adiabatic hypothesis
which states that 
\begin{eqnarray*}
\lim_{t\rightarrow -\infty} |\Phi_n(t)\rangle
&=& |\Phi_0\rangle,
\end{eqnarray*}
so that
\begin{eqnarray*}
\lim_{t\rightarrow -\infty} V(t)
&=& 1,
\end{eqnarray*}
and $V(t)=U(t,-\infty)$.
Thus, $V(t)$ is unitary. This has two important consequences:
(i) the states $|\Phi_n(t)\rangle$ are complete 
at all times:
\begin{eqnarray*}
\sum_n |\Phi_n(t)\rangle \langle \Phi_n(t)|
&=&
V(t) \big(\sum_n |\Phi^0_n\rangle \langle \Phi^0_n| \big) V^\dagger(t)
=V(t)V^\dagger(t)=1,
\end{eqnarray*}
and (ii) the scalar products are conserved:
$\langle \Phi_m(t)|\Phi_n(t)\rangle=\langle \Phi^0_m|\Phi^0_n\rangle$.

To complete this section, we define the notion
of anti-time-ordering operator.
For any $X$ which can be written as a product of
field operators, the anti-time-ordering 
of $X$ is defined as $\barT(X)={\big(T(X^\dagger)\big)}^\dagger$
(see \cite{BrouderQG}). Notice that $\barT$ is linear
and its arguments commute. To understand the
physical meaning of $\barT$, we take an example.
If $t>t'$ we have
$\barT(A(t)B(t'))={\big(T(B^\dagger(t')A^\dagger(t))\big)}^\dagger
= {\big(A^\dagger(t)B^\dagger(t')\big)}^\dagger
= B(t')A(t)$.
Analogously, 
$\barT(A(t)B(t'))=A(t)B(t')$ if $t<t'$.
In other words, $\barT$ orders its arguments
so that the operators are on the right
when they occur later. 
This is true for any
number of arguments, and $\barT$ orders its
arguments in the reverse order with respect to
$T$. This is why $\barT$ is called the
anti-time-ordering operator.
The main example is
\begin{eqnarray*}
\barT\big(\psi(x)\barpsi(y)\big)
&=&
\theta(y^0-x^0)\psi(x)\barpsi(y)
-\theta(x^0-y^0) \barpsi(y)\psi(x).
\end{eqnarray*}

The most important application of the 
anti-time-ordering operator is the calculation
of $U^\dagger(t,t')$.
\begin{eqnarray}
U^\dagger(t,t') &=& 
{\Big(T \exp\big(-i\int_{t'}^t H^\inter(\tau) \dd \tau\big)
\Big)}^\dagger
\nonumber\\&=&
\barT \exp\big(i\int_{t'}^t H^\inter(\tau) \dd \tau\big),
\label{Udagger(tt')}
\end{eqnarray}
because $H^\inter(\tau)$ is Hermitian.

\subsection{Evolution of density matrices}
If $|\Phi^0_n(t)\rangle$ are solutions of the
Schr\"odinger equation for $H_0$, a density
matrix $\hat\rho^0_S(t)$ in the Schr\"odinger
picture has the following general form
\begin{eqnarray*}
\hat\rho^0_S(t) &=&
\sum_{mn} \rho_{nm} |\Phi^0_n(t)\rangle\langle\Phi^0_m(t)|,
\end{eqnarray*}
where $\rho_{nm}$ is a Hermitian matrix with 
non-negative eigenvalues such that $\sum_n \rho_{nn}=1$.
For later convenience, we do not require $\rho_{nm}$
to be a diagonal matrix.
From the Schr\"odinger equation, we see that
the density matrix satisfies the equation
\begin{eqnarray*}
\frac{\partial \hat\rho^0_S(t)}{\partial t}
&=& -i[H_0,\hat\rho^0_S(t)].
\end{eqnarray*}

As for the wavefunctions, we define the density
matrix in the Heisenberg representation  $\hat\rho$
by $\hat\rho=\ee^{iH_0t}\hat\rho^0_S(t)\ee^{-iH_0t}= 
\sum_{mn} \rho_{nm} |\Phi^0_n\rangle \langle\Phi^0_m|$.
Notice that $\hat\rho$ does not depend on time.

In the interacting case, we look for a density
matrix $\hat\rho_S(t)$ in the Schr\"odinger picture
that we write
\begin{eqnarray*}
\hat\rho_S(t) &=&
\sum_{mn} \rho_{nm} |\Phi^S_n(t)\rangle\langle\Phi^S_m(t)|.
\end{eqnarray*}
It satisfies the equation 
\begin{eqnarray*}
\frac{\partial \hat\rho_S(t)}{\partial t}
&=& -i[H,\hat\rho_S(t)].
\end{eqnarray*}
We go to the interaction picture by defining
$\hat\rho_I(t)=\ee^{iH_0t}\hat\rho_S(t)\ee^{-iH_0t}$,
which satisfies the equation
\begin{eqnarray*}
\frac{\partial \hat\rho_I(t)}{\partial t}
&=& -i[H^\inter(t),\hat\rho_I(t)].
\end{eqnarray*}
Now it is easy to see that the density matrix
\begin{eqnarray*}
\hat\rho_I(t) &=&
\sum_{mn} \rho_{nm} V(t)|\Phi^0_n\rangle\langle\Phi^0_m|V^\dagger(t)
= \sum_{mn} \rho_{nm}|\Phi_n(t)\rangle\langle\Phi_m(t)|,
\end{eqnarray*}
satisfies the above equation. In other words, the
density matrix $\hat\rho_I(t)$ describes the interacting system
and it can be considered as the interacting density
matrix that evolved from the non-interacting density matrix
$\hat\rho$ because of the interactions.

\subsection{Evolution of expectation values}
\label{evolutionev}
The value of the observable $A(t)$ (in the interaction
picture) for a system in a mixed state described by the
 density matrix $\hat\rho_I$ is (see \cite{Bohm} p.314)
\begin{eqnarray}
\langle A(t)\rangle &=& \tr\big(\hat\rho_I A(t)\big)=
\sum_{mn} \rho_{mn}\langle \Phi_m(t)| A(t) | \Phi_n(t)\rangle
\nonumber\\
&=&
\sum_{mn} \rho_{mn}\langle \Phi^0_m| V^\dagger(t) 
A(t) V(t) | \Phi^0_n\rangle
=\tr\big(\hat\rho  V^\dagger(t) A(t) V(t) \big)
\nonumber\\&=&
\tr\big(\hat\rho   U(-\infty,t) A(t) U(t,-\infty)  \big).
\label{ArhoU}
\end{eqnarray}
The group property of $U(t,t')$ enables us to derive
\begin{eqnarray}
\langle A(t)\rangle &=& 
\tr\big(\hat\rho U(-\infty,t) U(t,+\infty)U(+\infty,t)A(t) 
U(t,-\infty)\big)
\nonumber\\&=&
\tr\big(\hat\rho U(-\infty,+\infty) U(+\infty,t)A(t) 
U(t,-\infty) \big),
\nonumber\\&=&
\tr\big(\hat\rho S^{-1} T(A(t)\ee^{-i\action})\big),
\label{ArhoS}
\end{eqnarray}
where the interacting action is (up to a sign)
$\action=\int_{-\infty}^\infty  H^\inter(\tau) \dd \tau$
and where the S-matrix is defined by
$S=U(+\infty,-\infty)=T(\ee^{-i\action})$.
The last line of (\ref{ArhoS})
was derived as follows. By equation (\ref{U(tt')})
\begin{eqnarray*}
U(+\infty,t)A(t) U(t,-\infty) &=&
T \big(\exp(-i\int_{t}^\infty H^\inter(\tau) \dd \tau)\big)
A(t)
\\&&
T \big(\exp(-i\int_{-\infty}^t H^\inter(\tau) \dd \tau)\big).
\end{eqnarray*}
In that expression, the operators are on the
left when their time arguments are larger. Thus, they
are time ordered and we can rewrite this
\begin{eqnarray*}
U(+\infty,t)A(t) U(t,-\infty) &=&
T \Big(\exp(-i\int_{t}^\infty H^\inter(\tau) \dd \tau)
A(t) 
\\&&
\exp(-i\int_{-\infty}^t H^\inter(\tau) \dd \tau)\Big).
\end{eqnarray*}
The arguments of the time-ordering operator commute, thus
\begin{eqnarray*}
U(+\infty,t)A(t) U(t,-\infty) &=&
T \Big(A(t)
\exp(-i\int_{t}^\infty H^\inter(\tau) \dd \tau)
\\&&
\exp(-i\int_{-\infty}^t H^\inter(\tau) \dd \tau)\Big)
\\&=&
T \Big(A(t)
\exp(-i\int_{-\infty}^\infty H^\inter(\tau) \dd \tau)\Big)
\\&=&
T \big(A(t) \ee^{-i\action}\big).
\end{eqnarray*}

To obtain equation (\ref{ArhoS}), we inserted
$1=U(t,+\infty)U(+\infty,t)$ before $A(t)$ in 
equation (\ref{ArhoU}). Of course, we can also
insert $1=U(t,+\infty)U(+\infty,t)$ after $A(t)$ in
equation (\ref{ArhoU}). This gives us the
alternative formula
\begin{eqnarray}
\langle A(t)\rangle &=& 
\tr\big(\hat\rho T^*(A(t)\ee^{i\action})S \big).
\label{ArhoSm1}
\end{eqnarray}

\subsection{Correlation functions}
Finally, we shall have to determine the correlation
function between an observable $A(t)$ at time $t$
and an observable $B(t')$ at time $t'$.
To do this, we must determine which picture
must be used to describe the observables at
two different times.
It turns out that the Heisenberg picture
does the job. There are three reasons for this:
(i) the equation for the observables in the Heisenberg
picture are similar to the equations for the
corresponding classical observables (see \cite{Bohm} p.316),
(ii) the correlation functions of observables calculated in the
Heisenberg picture agree with the experimental measurement
of these observables (see \cite{MandelWolf}, p. 655),
(iii) the quantum description of photodetectors shows
that they measure the correlation functions of the
photon field in the Heisenberg picture
(see \cite{MandelWolf}, chapter 14).

The relation between the observables in the
Schr\"odinger and Heisenberg pictures is given
by $A_H(t)=V_S^\dagger(t) A_S(t) V_S(t)$ 
(see \cite{Gross}, p. 143), where
$V_S$ satisfies the Schr\"odinger equation for
the full Hamiltonian $H_S(t)$:
\begin{eqnarray*}
\frac{\partial V_S(t)}{\partial t}
&=& -iH_S(t)V_S(t).
\end{eqnarray*}
The standard boundary condition is $V_S(0)=1$
and the solution of this equation is
$V_S(t)=\ee^{-iH_0t} U(t,0)$.
The boundary condition means that the Heisenberg and Schr\"odinger
pictures coincide at $t=0$. Therefore, the 
time-independent density matrix of the Heisenberg
picture is equal to the Schr\"odinger
density matrix at $t=0$, i.e. 
\begin{eqnarray*}
\hat\rho_H &=& \hat\rho_S(0)=
\sum_{mn} \rho_{nm} |\Phi^S_n(0)\rangle\langle\Phi^S_m(0)|.
\end{eqnarray*}
The correlation function for the two variables
$A(t)$ and $B(t')$ is now
\begin{eqnarray*}
\langle A(t) B(t')\rangle
&=& 
\tr\big(\hat\rho_H A_H(t) B_H(t')\big)
\\&=&
\sum_{mn} \rho_{nm} \langle\Phi^S_m(0)| A_H(t)
B_H(t')|\Phi^S_n(0)\rangle
\\&=&
\sum_{mn} \rho_{nm} \langle\Phi^0_m| U(-\infty,0)A_H(t)
B_H(t')U(0,-\infty)|\Phi^0_n\rangle
\\&=&
\sum_{mn} \rho_{nm} \langle\Phi^0_m| U(-\infty,0)
U(0,t) \ee^{iH_0 t}A_S(t)\ee^{-iH_0 t}U(t,0)
\\&&
U(0,t') \ee^{iH_0 t'}B_S(t')\ee^{-iH_0 t'}U(t',0)
U(0,-\infty)|\Phi^0_n\rangle
\\&=&
\tr\big( \hat\rho U(-\infty,t) A(t) U(t,t') B(t') U(t',-\infty)\big).
\end{eqnarray*}

As in the previous subsection, the group property of the
evolution operators $U(t,t')$ enables us to rewrite
three kinds of correlation functions, for the
operator product of fields, the time-ordered product of fields
and the anti-time-ordered  product of fields.
\begin{eqnarray*}
\langle A(t) B(t')\rangle &=&
\tr\big( \hat\rho T^*(A(t) \ee^{i\action})
         T(B(t') \ee^{-i\action})\big),\\
\langle T(A(t) B(t'))\rangle &=&
\tr\big( \hat\rho S^{-1}
         T(A(t) B(t') \ee^{-i\action})\big),\\
\langle \barT(A(t) B(t'))\rangle &=&
\tr\big( \hat\rho T^*(A(t) B(t') \ee^{i\action})
         S\big).
\end{eqnarray*}

\section{Functional derivative approach}
\subsection{Functional derivatives of the S-matrix}
\label{fundersect}
The use of functional derivatives in quantum
field theory was advocated by Schwinger \cite{Schwinger}.
The S-matrix for a nonrelativistic systems of electrons
with Coulomb interaction is given by
\begin{eqnarray*}
S &=& U(+\infty,-\infty) = T (\ee^{-i\action}).
\end{eqnarray*}
In solid-state physics, we usually consider the
free and interaction Hamiltonians (\cite{Abrikosov} p.44)
\begin{eqnarray*}
H_0 &=& \sum_{s=1}^2 \int \barpsi_s(t,\bfr) (-\frac{\Delta}{2m} + U_N(\bfr))
   \psi_s(t,\bfr) \dd\bfr,\\
H^\inter(t) &=&  \frac{1}{2} \sum_{s,s'} \int
   \barpsi_s(t,\bfr)\barpsi_{s'}(t,\bfr')  V_e(\bfr-\bfr') 
   \psi_{s'}(t,\bfr')\psi_s(t,\bfr) \dd\bfr \dd\bfr',
\end{eqnarray*}
where $U_N(\bfr)$ describes the interaction with the nuclei
and $V_e(\bfr)=e^2/(4\pi\epsilon_0 |\bfr|)$ the electron-electron
interaction.
We define now an S-matrix which depends on
two external fermion sources $\eta(x)$ and
$\bareta(x)$ as
\begin{eqnarray*}
S(\bareta,\eta) &=& 
T \exp\big(-i\action +i\int \bareta(x)\psi(x) \dd x
  +i \int \barpsi(x)\eta(x) \dd x \big).
\end{eqnarray*}
For a nonrelativistic fermion, $\psi(x)$ and 
$\barpsi(x)$ are two-component vectors. Thus,
the sources are also two-component vectors
and
\begin{eqnarray*}
\bareta(x)\psi(x) &=& \sum_{s=1}^2
  \bareta_s(x)\psi_s(x),
\quad\quad
\barpsi(x)\eta(x) = \sum_{s=1}^2
  \barpsi_s(x)\eta_s(x).
\end{eqnarray*}

The functional derivative 
with respect to the fermion source $\eta(x)$
satisfies
\begin{eqnarray}
\frac{\delta}{\delta\eta(x)}\eta(y) &=& \delta(x-y),
\quad
\frac{\delta}{\delta\eta(x)}\bareta(y) = 0,
\nonumber\\
\frac{\delta}{\delta\eta(x)} (uv) &=& \big(\frac{\delta u}
     {\delta\eta(x)}\big)v
+(-1)^{|u|} u \big(\frac{\delta v}{\delta\eta(x)}\big).
\label{Leibniz}
\end{eqnarray}
In this equation, we assumed that $u$ is the product of
a certain number of fermion fields or sources, and this
number is denoted by $|u|$.
Similar relations
are satisfied by the functional derivative with respect
to $\bareta(x)$. Equation (\ref{Leibniz}) is known as 
Leibniz' rule.

The sources $\eta$ and $\bareta$ anticommute, so 
the functional derivatives anticommute:
\begin{eqnarray*}
\frac{\delta^2}{\delta\eta(x)\delta\eta(y)} &=& 
-\frac{\delta^2}{\delta\eta(y)\delta\eta(x)}.
\end{eqnarray*}

To see how functional derivatives act with
respect to the time-ordering operator, we
first notice that the sources
can be taken out of the time-ordering operator.
For example, if $x^0>y^0$ 
\begin{eqnarray*}
T(\bareta(x)\psi(x)\barpsi(y)\eta(y))
&=& \bareta(x)\psi(x)\barpsi(y)\eta(y) =
\bareta(x)\eta(y) \psi(x)\barpsi(y)
\\&=& \bareta(x)\eta(y) T(\psi(x)\barpsi(y)),
\end{eqnarray*}
if $x^0<y^0$ 
\begin{eqnarray*}
T(\bareta(x)\psi(x)\barpsi(y)\eta(y))
&=& \barpsi(y)\eta(y) \bareta(x)\psi(x)=
-\bareta(x)\eta(y) \barpsi(y)\psi(x)
\\&=& \bareta(x)\eta(y) T(\psi(x)\barpsi(y)).
\end{eqnarray*}
Thus, the functional derivative with respect to
$\eta(x)$ or $\bareta(x)$ commutes with the time-ordering
operator.
In particular
\begin{eqnarray*}
\frac{\delta S(\bareta,\eta)}{\delta\bareta(x)}|_{\bareta=\eta=0} &=&
i T \big(\psi(x)\ee^{-i\action}\big)
 =
i U(+\infty,t) \psi(x) U(t,-\infty),\\
\frac{\delta S(\bareta,\eta)}{\delta\eta(x)}|_{\bareta=\eta=0} &=&
-i T \big(\barpsi(x)\ee^{-i\action}\big)
= -i U(+\infty,t) \barpsi(x) U(t,-\infty),
\end{eqnarray*}
where $x=(t,\bfr)$ \cite{Itzykson} and the minus sign in the
last equation comes from the fact that the functional derivative
must jump over $\barpsi(x)$ to reach $\eta(x)$
in the definition of $S(\bareta,\eta)$.
With this definition, we can write
\begin{eqnarray}
\langle \psi_H(x)\rangle &=& 
i\frac{\delta}{\delta\bareta(x)} \sum_{mn} \rho_{nm} 
\langle\Phi_0^m|S(\bareta,\eta)^{-1} S(\bareta,\eta)|\Psi_0^n\rangle|_{\bareta=\eta=0}.
\label{rangle_rho}
\end{eqnarray}
In the vacuum, the density matrix is $|0\rangle\langle 0|$
and 
\begin{eqnarray*}
\langle \psi_H(x)\rangle_0 &=&
i\frac{\delta}{\delta\bareta(x)} 
\langle 0|S(0,0)^{-1}
S(\bareta,\eta)|0\rangle|_{\bareta=\eta=0}.
\end{eqnarray*}
One then invokes the ``stability of the vacuum'' \cite{Itzykson}
to derive
\begin{eqnarray}
\langle \psi_H(x)\rangle_0 &=&
i\frac{\delta}{\delta\bareta(x)}
\langle 0|S(0,0)^{-1}|0\rangle
\langle 0|S(\bareta,\eta)|0\rangle|_{\bareta=\eta=0}
\nonumber\\&=&
i\frac{\delta}{\delta\bareta(x)}
\frac{\langle 0|S(\bareta,\eta)|0\rangle}
{\langle 0|S(0,0)|0\rangle}
|_{\bareta=\eta=0},
\label{GellMannLow}
\end{eqnarray}
which is the Gell-Mann and Low formula \cite{GellMann}.
The denominator is a pure phase, thus the
main problem is to calculate the numerator of
equation (\ref{GellMannLow}).
A standard result of the functional derivative approach 
\cite{Itzykson,Redmond} is that
the interacting S-matrix $S(\bareta,\eta)$ can
be obtained from the non-interacting S-matrix $S^0(\bareta,\eta)$
with
$S^0(\bareta,\eta) =
T \exp\big(i \int \bareta(x)\psi(x) + \barpsi(x)\eta(x)\dd x \big)$
by the equation
\begin{eqnarray*}
S(\bareta,\eta) &=& 
\exp\big(-i\int_{-\infty}^\infty 
     H^\inter(\frac{-i\delta}{\delta\bareta(x)},
               \frac{i\delta}{\delta\eta(x)}) \dd t\big)
               S^0(\bareta,\eta),
\end{eqnarray*}
where $x=(t,\bfr)$.
For a state described by a density matrix
$\hat\rho=\rho_{nm} |\Psi_0^n\rangle\langle\Phi_0^m|$,
the Gell-Mann and Low formula does not hold
and we must deal with the term $S(\bareta,\eta)^{-1}$
in equation (\ref{rangle_rho}). This is done
by doubling the sources.

\subsection{Source doubling}
The idea of doubling the sources was proposed independently
by Schwinger \cite{SchwingerJMP} and
Symanzik \cite{Symanzik,Symanzik2}.
It is a basic technique of nonequilibrium quantum
field theory 
\cite{DuBois,Hall,Landsman,Chou,Rammer,Henning,FauserWolter2}
where it is also known as the closed time-path
Green function formalism.
For equilibrium quantum field theory, Wagner showed
that it can be useful to triple the sources
\cite{Wagner}.
In equation (\ref{rangle_rho}), we have the
operator product of $S(\bareta,\eta)^{-1}$ and
$S(\bareta,\eta)$. We cannot obtain an operator
product by functional derivatives, because
they generate time-ordered products  of operators.
Therefore, we shall use sources to calculate
$S(\bareta,\eta)^{-1}$ and sources to calculate $S(\bareta,\eta)$:
we define
\begin{eqnarray}
Z_\rho &=&
\sum_{mn} \rho_{nm} 
\langle\Phi_0^m|S(\bareta_-,\eta_-)^{-1}
S(\bareta_+,\eta_+)|\Psi_0^n\rangle.
\label{defZrho}
\end{eqnarray}
Here $Z_\rho$ is a function of the sources
$\bareta_-,\eta_-,\bareta_+,\eta_+$.
Notice that  $Z_\rho=1$ when $\bareta_-=\bareta_+$
and $\eta_-=\eta_+$, because
$S(\bareta,\eta)^{-1} S(\bareta,\eta)=1$
and $\tr\hat\rho=1$.
To calculate $S(\bareta,\eta)^{-1}$, we recall
that $S$ is unitary, so that
\begin{eqnarray*}
S(\bareta_-,\eta_-)^{-1} &=&
S(\bareta_-,\eta_-)^\dagger
\\ &=& 
\Big(T \exp\big(-i\action
  +i\int \bareta_-(x)\psi(x)
  + \barpsi(x)\eta_-(x) \dd x \big)\Big)^\dagger
\\&=&
\barT \exp\big(i\action
  -i\int \bareta_-(x)\psi(x) +
             \barpsi(x)\eta_-(x) \dd x \big),
\end{eqnarray*}
where $\barT$ is the anti-time-ordering operator
first considered by Dyson 
\cite{Dyson51I,Dyson51II} (see also \cite{Pauli} p.94),
which orders operators according to decreasing times.
For example,
\begin{eqnarray*}
\barT\big(\psi(x)\barpsi(y)\big)
&=&
\theta(y^0-x^0)\psi(x)\barpsi(y)
-\theta(x^0-y^0) \barpsi(y)\psi(x).
\end{eqnarray*}
As for $S(\bareta_+,\eta_+)$, we can write
\begin{eqnarray*}
S(\bareta_-,\eta_-)^{-1} &=& 
\exp\big(i\int_{-\infty}^\infty 
     H^\inter(\frac{i\delta}{\delta\bareta_-(x)},
               \frac{-i\delta}{\delta\eta_-(x)}) \dd t\big)
               S^0(\bareta,\eta)^{-1},
\end{eqnarray*}
where $x=(t,\bfr)$  and
\begin{eqnarray*}
S^0(\bareta_-,\eta_-)^{-1} &=&
\barT \exp\big( -i\int \bareta_-(x)\psi(x) \dd x 
         -i \int \barpsi(x)\eta_-(x) \dd x \big).
\end{eqnarray*}
If we put all this together, we obtain
\begin{eqnarray}
Z_\rho &=&
\ee^{-iD} Z^0_\rho,
\label{Zrho=DZ0rho}
\end{eqnarray}
where
\begin{eqnarray}
D &=&
\int_{-\infty}^\infty
     H^\inter(\frac{i\delta}{\delta\bareta_+(x)},
               \frac{-i\delta}{\delta\eta_+(x)})
- H^\inter(\frac{-i\delta}{\delta\bareta_-(x)},
               \frac{i\delta}{\delta\eta_-(x)}) \dd t
\label{defh}
\end{eqnarray}
and
\begin{eqnarray}
Z^0_\rho &=&
\sum_{mn} \rho_{nm} 
\langle\Phi^0_m|S^0(\bareta_-,\eta_-)^{-1}
S^0(\bareta_+,\eta_+)|\Psi^0_n\rangle.
\label{defZrho0}
\end{eqnarray}
Notice that the functional derivatives with respect
to $\eta_-(x)$ and $\bareta_-(x)$ correspond to
anti-time-ordering.

These are the basic equations for the calculation
of $Z_\rho$. The next step is now the evaluation
of $Z^0_\rho$.

\section{Calculation of $Z^0_\rho$}
In the calculation of $Z^0_\rho$, we first
write $S^0(\bareta_-,\eta_-)^{-1}
 S^0(\bareta_+,\eta_+)$ in terms of normally
ordered operators, then we calculate the
trace of the normal ordered term.
The use of normal order is very convenient to
calculate matrix elements.

\subsection{Normal ordering}
If we call 
$A = -i\int \bareta_-(x)\psi(x)+\barpsi(x)\eta_-(x)\dd x$ 
and
$B = i\int \bareta_+(x)\psi(x)+\barpsi(x)\eta_+(x)\dd x$,
we have $S^0(\bareta_-,\eta_-)^{-1}S^0(\bareta_+,\eta_+)=\barT(\ee^A)T(\ee^B)$.
We want to write $\barT(\ee^A)T(\ee^B)$ as the product of
scalar terms with the normally ordered exponential ${:}\ee^{A+B}{:}$.
To achieve this, we use the identity giving
the time-ordered exponential in terms of the
normally-ordered exponential: $T(\ee^B) = \ee^\beta {:}\ee^B{:}$
(see eq.(4-73) p. 183 of ref. \cite{Itzykson}), where
\begin{eqnarray*}
\beta &=&
-\int \bareta_+(x) \langle 0|T\psi(x)\barpsi(y)|0\rangle
           \eta_+(y) \dd x \dd y.
\end{eqnarray*}
This identity
is a generating function for Wick's theorem.
The same proof leads to the corresponding identity for
the anti-time-ordered products
$\barT(\ee^A) = \ee^\alpha {:}\ee^A{:}$,
where
\begin{eqnarray*}
\alpha &=&
-\int \bareta_-(x) \langle
0|\barT\psi(x)\barpsi(y)|0\rangle
           \eta_-(y) \dd x \dd y.
\end{eqnarray*}
Thus, $\barT(\ee^A)T(\ee^B)=\ee^{\alpha+\beta} {:}\ee^A{:}{:}\ee^B{:}$
and it remains to normally order the operator product of
${:}\ee^A{:}$ and ${:}\ee^B{:}$.
To do that, we write the operator exponential in terms of a
normally ordered exponential
$\ee^A = \ee^{\alpha'} {:}\ee^A{:}$ and 
$\ee^B = \ee^{\beta'} {:}\ee^B{:}$,
where
\begin{eqnarray*}
\alpha' &=&
-\frac{1}{2}\int \bareta_-(x) \langle 0|[\psi(x),\barpsi(y)]|0\rangle
           \eta_-(y) \dd x \dd y,\\
\beta' &=&
-\frac{1}{2}\int \bareta_+(x) \langle 0|[\psi(x),\barpsi(y)]|0\rangle
           \eta_+(y)  \dd x \dd y.
\end{eqnarray*}
This identity is the generating function for Wick's
theorem for operator products.
To obtain this result
we start from eq.(4-72) p. 183 of ref. \cite{Itzykson}
and we use the fact that
$\{\psi^{(-)}(x),\barpsi^{(+)}(y)\}=\langle 0 | \barpsi(y)
  \psi(x)|0\rangle$ and
$\{\barpsi^{(-)}(x),\psi^{(+)}(y)\}=\langle 0 | \psi(y)
  \barpsi(x)|0\rangle$.
Thus, ${:}\ee^A{:}\,{:}\ee^B{:}=\ee^{-\alpha'-\beta'} \ee^A\ee^B$.
To transform the product $\ee^A\ee^B$, we
can employ the classical expression
$\ee^A \ee^B = \ee^{A+B+[A,B]/2}$, valid when
$[A,B]$ commutes with $A$ and $B$ (eq. (4-15)
p. 167 of ref. \cite{Itzykson}).
This is the case here because
\begin{eqnarray*}
[A,B] &=& \int \bareta_-(x) \{\psi(x),\barpsi(y)\} \eta_+(y)
\\&&
+\eta_-(x) \{\barpsi(x),\psi(y)\} \bareta_+(y)
           \dd x \dd y,
\end{eqnarray*}
is not an operator but a function
(i.e. $\{\psi(x),\barpsi(y)\}=
\langle 0|\{\psi(x),\barpsi(y)\} |0\rangle$.
Now, we transform again the exponential 
$\ee^{A+B}$ into a normally ordered exponential
by
$\ee^{A+B} = \ee^{\gamma'} {:}\ee^{A+B}{:}$, where
\begin{eqnarray*}
\gamma' &=&
-\frac{1}{2}\int \bareta_d(x) \langle 0|[\psi(x),\barpsi(y)]|0\rangle
           \eta_d(y) \dd x \dd y,
\end{eqnarray*}
with $\bareta_d=\bareta_+-\bareta_-$
and $\eta_d=\eta_+-\eta_-$.
Putting all this together, we find
${:}\ee^A{:} {:}\ee^B{:} = \ee^\gamma {:}\ee^{A+B}{:}$,
with $\gamma=-\alpha'-\beta'+[A,B]/2+\gamma'$, so that
\begin{eqnarray*}
\gamma &=&
\int \eta_-(x)
\langle 0|\barpsi(x)\psi(y)|0\rangle \bareta_+(y)
\nonumber\\&&
+\bareta_-(x) \langle 0|\psi(x)\barpsi(y) |0\rangle
\eta_+(y) \dd x \dd y.
\end{eqnarray*}
Thus, $\barT(\ee^A)T(\ee^B)=\ee^{\alpha+\beta+\gamma} {:}\ee^{A+B}{:}$.
The calculation of $\alpha+\beta+\gamma$ gives us 
\begin{eqnarray*}
\barT(\ee^A)T(\ee^B) &=& \exp[-i\int \bareta(x) G^0_0(x,y)\eta(y)\dd x\dd y]
N^0(\bareta_d,\eta_d).
\end{eqnarray*}
The two-dimensional vectors $\eta$ and $\bareta$ are
\begin{eqnarray*}
\eta(x) = \left( \begin{array}{c}
         \eta_+(x) \\
         \eta_-(x)
         \end{array}\right)
\qquad
\bareta(x) = \left( \begin{array}{c}
         \bareta_+(x) \\
         \bareta_-(x)
         \end{array}\right),
\end{eqnarray*}
the free Green function is
\begin{eqnarray}
G^0_0(x,y) &=& \left( \begin{array}{cc}
    -i\langle 0 | T\big(\psi(x)\barpsi(y)\big)|0\rangle
       & -i\langle 0 | \barpsi(y)\psi(x)|0\rangle \\
   i\langle 0 | \psi(x)\barpsi(y)|0\rangle
        & -i\langle 0 |
\barT\big(\psi(x)\barpsi(y)\big)|0\rangle
         \end{array}\right),
\label{freeGreen}
\end{eqnarray}
and the normally ordered exponential is
\begin{eqnarray*}
N^0(\bareta_d,\eta_d) &=&
{:}\exp\big[ i\int \bareta_d(x)\psi(x)+\barpsi(x)\eta_d(x)\dd x\big]{:}.
\end{eqnarray*}
Notice that the Green function is a solution of the equation
\begin{eqnarray*}
\big(i\partial_{x^0} -h_0(x)\big)
   G^0_0(x,y) = \left( \begin{array}{cc}
    \delta(x-y) &  0  \\
   0 & -\delta(x-y)
      \end{array}\right).
\end{eqnarray*}

Finally, the generating function is
\begin{eqnarray*}
Z^0_\rho&=& \exp[-i\int \bareta(x) G^0_0(x,y)\eta(y)\dd x\dd y]
\tr[\hat\rho N^0(\bareta_d,\eta_d)].
\end{eqnarray*}
A similar expression is given in ref. \cite{Chou}.

Schwinger \cite{SchwingerJMP} 
showed that this expression can be
rewritten in terms of advanced and retarded
Green functions, using the sources
$\bareta_m=(\bareta_+ + \bareta_-)/2$
and
$\eta_m=(\eta_+ + \eta_-)/2$.
\begin{eqnarray*}
Z^0_\rho&=& \exp[-\int \bareta_d(x) G^0_r(x,y) \eta_m(y)
 - \bareta_m(x) G^0_a(x,y) \eta_d(y)
\\&&
 + \frac{1}{2} \bareta_d(x) G^0_c(x,y) \eta_d(y)
 \dd x \dd y]
\tr[\hat\rho N^0(\bareta_d,\eta_d)],
\end{eqnarray*}
with
\begin{eqnarray*}
 G^0_r(x,y) &=&  \theta(x^0-y^0) \langle 0
\{\psi(x),\barpsi(y)\}|0\rangle,\\
 G^0_a(x,y) &=& -\theta(y^0-x^0) \langle 0 |
\{\psi(x),\barpsi(y)\}|0\rangle,\\
 G^0_c(x,y) &=& \langle 0 | [\psi(x),\barpsi(y)] |0\rangle.
\end{eqnarray*}

\subsection{Calculation of $\tr[\hat\rho N^0(\bareta_d,\eta_d)]$}
\label{unsection}
The calculation of $\tr[\hat\rho N^0(\bareta_d,\eta_d)]$
is relegated to appendix because it is rather technical.
We give here the results.
The unperturbed eigenstates of $H_0$ will now be called
$|K\rangle$ and $|L\rangle$ instead of 
$|\Phi^0_m\rangle$ and $|\Phi^0_n\rangle$.
They are defined from the vacuum $|0\rangle$ by
application of creation operators
$|K\rangle = b^\dagger_{i_N}\dots b^\dagger_{i_1} |0\rangle$
and
$|L\rangle = b^\dagger_{j_N}\dots b^\dagger_{j_1} |0\rangle$.
Here, $N$ is the number of electrons and
the indices $i_k$ and $j_k$ take their values in the
set of indices of the $M$ orbitals. 
We assume that the indices are ordered:
$i_1<\dots<i_N$ and $j_1<\dots<j_N$.
If we take the example of Cr$^{3+}$, the number of $d$
electrons is $N=3$ and the number of $d$ orbitals is
$M=10$.  We assume that the orbitals are ordered in such a way
that the $M$ orbitals that come into play are numbered
from $n=1$ to $n=M$.
We define integrals of the product of 
the wavefunctions with external sources
by $\bara_n=\int \bareta_d(x)u_n(x)\dd x$
and $\alpha_n=\int \baru_n(x)\eta_d(x)\dd x$,
where $u_n(x)=\ee^{-i\epsilon_n t} u_n(\bfr)$
and $\baru_n(x)=\ee^{i\epsilon_n t} \baru_n(\bfr)$,
with $x=(t,\bfr)$.
Recall that $\bareta_d=\bareta_+-\bareta_-$ and $\eta_d=\eta_+-\eta_-$.
The result can now be stated in its simplest form
as $\tr\big(\hat\rho N^0(\bareta_d,\eta_d)\big)=
\sum_{KL} \rho_{LK} N^0_{KL}$ with
\begin{eqnarray}
N^0_{KL} &=&
  \langle K | N^0(\bareta_d,\eta_d)|L\rangle\nonumber\\
&=&
\exp\Big(\sum_{n=1}^M \frac{\partial^2}{\partial \alpha_n
      \partial \bara_n}\Big)
\bara_{j_1} \alpha_{i_1}
\dots
\bara_{j_N} \alpha_{i_N}.
\label{alternativex}
\end{eqnarray}
A more explicit but more cumbersome form of this result
is given in the appendix.

It is interesting to consider the particular
case of a closed shell (see equation (\ref{closedshell})
in the appendix).
This happens when
all orbitals are occupied, i.e. $N=M$. Then there
is only one state, $\hat\rho=1$ and 
\begin{eqnarray*}
\tr\big(\hat\rho N^0\big) &=&
\prod_{k=1}^N (1+\bara_{i_k}\alpha_{i_k}).
\end{eqnarray*}

\section{The Hopf algebra of derivations}
The term quantum group has a broad meaning \cite{Majid},
ranging from general Hopf algebras to $q$-deformed groups.
In this section we use the more precise term of Hopf algebra.

We introduce now the Hopf algebra of functional derivations
$\calD$, which plays a vital role in this paper.
In particular, the calculation of
$\tr(\hat\rho N^0(\bareta_d,\eta_d))$ and 
the resummation leading to the hierarchy
of Green functions for degenerate systems
make essential use of the Hopf structure
of $\calD$. Writing this hierarchy 
without Hopf-algebraic tools would
be quite cumbersome.
Since the introduction of the Hopf algebra
of renormalization by Kreimer \cite{Kreimer98},
it has become clear that Hopf algebras are
going to play a substantial role in quantum field
theory \cite{BrouderOecklI,BrouderQG}.

Many textbooks on Hopf algebras are now
available \cite{Kassel,Majid} but we shall
use only a very limited amount of this theory.
For the convenience of the reader, we give
now a short survey of the Hopf algebra
of derivations.

\subsection{A familiar example of coproduct}
The most unusual object of a Hopf algebra is
the coproduct. To make the reader familiar with
this concept, we present it in the case of 
the algebra $\calA$ of differential operators
with constant coefficients.
We consider the coordinates
$x_1,\dots, x_n$ of an $n$-dimensional space,
and the differential operators
$P=\sum_{\boldsymbol{\alpha}} a_{\boldsymbol{\alpha}} D^{\boldsymbol{\alpha}}$,
where $\boldsymbol{\alpha}=(\alpha_1,\dots,\alpha_n)$ is a multi-index,
$a_{\boldsymbol{\alpha}}$ is a complex number and
$D^{\boldsymbol{\alpha}} = \partial_{1}^{\alpha_1}\dots \partial_{n}^{\alpha_1}$,
where $\partial_i$ denotes the partial derivative
$\partial/\partial_{x_i}$. It is clear that
$\calA$ is a vector space with basis $D^{\boldsymbol{\alpha}}$, where
$\boldsymbol{\alpha}$ runs over all the possible multi-indices.
$\calA$ is also an associative algebra with the product induced by the 
product of the basis elements 
$D^{\boldsymbol{\alpha}}D^{\boldsymbol{\beta}}=D^{\boldsymbol{\alpha}+\boldsymbol{\beta}}$.
To this algebra we add a unit $\un$ such that $D\un=\un D=D$ for any element
$D$ of $\calA$.

In this context, the coproduct comes from the action of
a differential operator on a product of two functions.
The action of $\partial_i$ on the product $fg$ is given
by the Leibniz rule
$\partial_i(fg)=(\partial_i f) g + f (\partial_i g)$.
For a product ot two partial derivatives we have
\begin{eqnarray}
\partial_i\partial_j(fg) &=&
(\partial_i\partial_j f) g + f (\partial_i\partial_j g)
+ (\partial_i f) (\partial_j g)
+ (\partial_j f) (\partial_i g).
\label{parpar}
\end{eqnarray}
More generally, for any differential operator $P\in\calA$,
we can write $P(fg)$ as a sum of terms that are
the product of a differential operator acting on $f$
and a differential operator acting on $g$.
We write this using Sweedler's notation
$P(fg)=\sum (P\i1 f) (P\i2 g)$.
For example, if $P=\partial_i$ we have
a sum of two terms, in the first term
$P\i1=\partial_i$ and $P\i2=\un$ (with the convention
that, for any function $f$, $\un f=f$)
and in the second term
$P\i1=\un$ and $P\i2=\partial_i$.
The idea of the coproduct is now to remove the reference
to the functions $f$ and $g$ and to keep only 
the sum of terms with $P\i1$ on the left and
$P\i2$ on the right.
This is done formally by defining the coproduct 
$\Delta$ from $\calA$ to $\calA\otimes\calA$ as
$\Delta P = \sum P\i1 \otimes P\i2$.
From the known properties of the action of 
a differential operator on a product of two functions
we deduce the following properties of the coproduct:
$\Delta \un = \un\otimes\un$,
$\Delta \partial_i = \partial_i\otimes \un + \un \otimes \partial_i$
and the recursive relation
\begin{eqnarray*}
\Delta (PP') &=& \sum (PP')\i1\otimes (PP')\i2
= \sum\sum P\i1 P\i1'\otimes P\i2 P\i2'.
\end{eqnarray*}
From the last rule we obtain
$\Delta (\partial_i\partial_j)=
(\partial_i\partial_j)\otimes\un + \un \otimes (\partial_i\partial_j)
+\partial_i\otimes\partial_j + \partial_j \otimes \partial_i$,
and we recover equation (\ref{parpar}).
The main property of the coproduct is its coassociativity, 
which means that 
\begin{eqnarray*}
\sum \Delta (P\i1) \otimes P\i2 &=&
\sum P\i1\otimes \Delta(P\i2) =
\sum P\i1\otimes P\i2 \otimes P\i3.
\end{eqnarray*}
For example, if $P=\partial_i$,
\begin{eqnarray*}
\sum P\i1\otimes P\i2 \otimes P\i3 &=&
\sum (\Delta \partial_i)\otimes \un + (\Delta \un)\otimes \partial_i
\\&=&
\partial_i\otimes \un\otimes \un + \un \otimes \partial_i \otimes \un
+ \un\otimes\un\otimes\partial_i.
\end{eqnarray*}

With this definition we can obtain the action of $P$
on a product of three functions
as
$P(fgh)=\sum (P\i1 f)\otimes (P\i2 g) \otimes (P\i3 h)$.

After this introduction, we can now define the
algebra of functional derivations.
The main changes are that the partial derivatives
are replaced by functional derivatives with
respect to external sources, and the fact
that the anticommutativity of external sources
generates signs in the formulas.

\subsection{The algebra structure of $\calD$}
The symbol $\partial$ is used to denote
the functional derivative with respect
to the external souces
$\eta(x)$ or $\bareta(x)$. More precisely,
since the external sources are two-dimensional
vectors, $\partial$ stands for the functional
derivative with respect to $\eta_s(x)$ or 
$\bareta_s(x)$, where $s=1$ or $s=2$.
Products of symbols stands for repeated derivations.
For instance, if
$\partial_1=\delta/\delta{\eta_1}(x)$,
$\partial_2=\delta/\delta{\bareta_2}(y)$ and
$\partial_3=\delta/\delta{\eta_2}(x)$,
then
\begin{eqnarray*}
\partial_1\partial_2\partial_3 &=&
\frac{\delta^3}{\delta\eta_1(x)\delta\bareta_2(y)\delta\eta_2(x)}.
\end{eqnarray*}
The functional derivatives anticommute, thus
$\partial\partial'=-\partial'\partial$ for any functional
derivatives $\partial$ and $\partial'$.
Therefore, for any functional derivative $\partial$,
we have $\partial\partial=0$.

A basis of the vector space $\calD$ of functional derivatives
with respect to external sources is given by the products
of derivations $\partial_1\dots\partial_n$ for all $n\ge 1$
and the unit $\un$. Here, the unit is not the constant function 1,
it is a symbol that satisfies $\un\partial=\partial \un=\partial$
for any functional derivative $\partial$.
Thus, for instance,
\begin{eqnarray*}
4 \un+2 \frac{\delta}{\delta\eta_1(x)}
+\frac{1}{6}
\frac{\delta^2}{\delta\bareta_2(y)\delta\eta_2(x)}           
\end{eqnarray*}
is an element of $\calD$.

In $\calD$, the terms of the form $\partial_1\dots\partial_n$
generate a subspace of $\calD$ denoted by $\calD_n$
(for $n>0$). The elements $\calD_0$ have the
form $\lambda \un$, where $\lambda$ is a complex number.
If $D\in\calD$ belongs to $\calD_n$ for some
$n$, we say that $D$ is homogeneous and its
degree, written $\degres(D)$, is $n$.
For instance $\degres(\un)=0$, $\degres(\partial)=1$,
$\degres(\partial\partial')=2$.
The vector space $\calD$ becomes an algebra
if we define the product of two elements of
$\calD$ to be the composition of derivations.
For instance, the product of 
$\partial_1$ and $\partial_2$ is $\partial_1\partial_2$.
This product is anticommutative. It can be checked
that $\calD$ is an associative algebra with unit $\un$.
Moreover, $\degres(DD')=\degres(D)+\degres(D')$
for any homogeneous elements $D$ and $D'$ of $\calD$.
From the degree $\degres(D)$ of a homogeneous element $D$
we can define its parity $|D|$ by $|D|=0$ if
$\degres(D)$ is even and $|D|=1$ if $\degres(D)$
is odd. If $|D|$=0 (resp. $|D|=1$) we say that
$D$ is even (resp. odd).

Now we prove a useful property of the product in $\calD$:
if $D$ and $D'$ are elements with a specific parity
$|D|$ and $|D'|$, then
\begin{eqnarray}
DD'=(-1)^{|D||D'|} D'D.
\label{DD'}
\end{eqnarray}
An important consequence of this is the fact that
an even element of $\calD$ commutes with all elements of $\calD$.
To prove equation (\ref{DD'}), we first show it
for homogeneous elements. We start with
$D=\partial$ and $D'=\partial'_1\dots\partial'_n$,
then 
$\partial\partial'_1\dots\partial'_n=
(-1)^n \partial'_1\dots\partial'_n\partial$
because $\partial$ must jump $n$ times over a
$\partial'$. Now, if $D=\partial_1\dots\partial_m$,
$\partial_m$ jumps over $D'$, giving $(-1)^n$,
then $\partial_{m-1}$ jumps over $D'$ giving another
$(-1)^n$, and so on until $\partial_1$ and we obtain
$DD'=(-1)^{mn}D'D=(-1)^{\degres(D)\degres(D')}$. 
Equation (\ref{DD'}) is recovered
because $(-1)^{\degres(D)\degres(D')}=(-1)^{|D||D'|}$.
If $D$ and $D'$ are not homogeneous but have a
definite parity, they can be written as sums of
homogeneous elements, and the result follows by
linearity.

\subsection{The coalgebra structure of $\calD$}
We introduce now the coproduct $\Delta$ of $\calD$. 
In concrete terms, the coproduct
of an element $D$ of $\calD$ is the sum of the
ways to split $D$ into the product of two elements
of $\calD$.
Formally, the coproduct is defined as a map
from $\calD$ to $\calD\otimes\calD$, where $\otimes$
stands for the tensor product.
We recall the main property of the tensor product \cite{Eisenbud}:
for any $D,D',E,E'\in \calD$ and $\lambda,\lambda',\mu,\mu'\in
\mathbb{C}$,
\begin{eqnarray*}
(\lambda D+\lambda' D')\otimes (\mu E + \mu' E')
&=& \lambda\mu D\otimes E + \lambda \mu' D\otimes E'
\\&&
+ \lambda'\mu D'\otimes E + \lambda' \mu' D'\otimes E'.
\end{eqnarray*}
The coproduct of the elements of smallest degrees is
given by
\begin{eqnarray}
\Delta \un &=& \un\otimes \un,\label{Delta1}\\
\Delta \partial &=& \partial\otimes \un + \un\otimes\partial.
\label{Deltapartial}
\end{eqnarray}
To define the coproduct of elements of higher degree, we
need a notation for the coproduct. Following Sweedler,
we write
$\Delta D=\sum D\i1\otimes D\i2$. For instance,
if $D=\partial$, the sum has two terms. The first
term is $D\i1=\partial$, $D\i2=\un$ the second term is
$D\i1=\un$, $D\i1=\partial$.
The coproduct can now be defined recursively by
\begin{eqnarray}
\Delta (D D') &=& \sum (-1)^{|D\i2||D\i1'|}
  (D\i1 D\i1') \otimes (D\i2 D\i2').
\label{DeltaDD'}
\end{eqnarray}
As an exercise, we calculate
$\Delta (\partial \partial')$,
so that $D=\partial$ and $D'=\partial'$.
Equation (\ref{Deltapartial}) gives us
$\Delta \partial = \partial\otimes \un + \un\otimes\partial$
and
$\Delta \partial' = \partial'\otimes \un + \un\otimes\partial'$.
The first term of $\Delta (\partial \partial')$
is obtained from formula (\ref{DeltaDD'}) with
$D\i1=\partial$,
$D\i2=\un$, $D\i1'=\partial'$
and $D\i2'=\un$. The degrees are $|D\i2|=0$,
$|D\i1'|=1$ and their product is $|D\i2||D\i1'|=0$
so we obtain the term
$\partial \partial'\otimes \un$. The other terms
are calculated analogously and the result is 
\begin{eqnarray*}
\Delta (\partial \partial') &=&  \partial \partial'\otimes \un
+  \un\otimes\partial \partial'
+ \partial\otimes \partial'
- \partial'\otimes\partial.
\end{eqnarray*}
The minus sign is due to the fact that the corresponding
term comes from $D\i1=\un$, $D\i2=\partial$,
$D\i1'=\partial'$ and $D\i2'=\un$, so that $|D\i2||D\i1'|=1$.

It can be checked \cite{Eisenbud} that the
coproduct of a basis element 
$D=\partial_1\dots\partial_n$ of $\calD$ is
\begin{eqnarray*}
\Delta D &=&
D\otimes \un + \un\otimes D
\nonumber\\&&+
\sum_{p=1}^{n-1} \sum_{\sigma} (-1)^\sigma
\partial_{\sigma(1)}\dots\partial_{\sigma(p)}
\otimes \partial_{\sigma(p+1)}\dots\partial_{\sigma(n)},
\end{eqnarray*}
where $\sigma$ runs over the $(p,n-p)$-shuffles
and $(-1)^\sigma$ is the signature of the
permutation $\sigma$.
Recall that a $(p,n-p)$-shuffle is a permutation $\sigma$ of
$\{1,\dots,n\}$ such that
$\sigma(1)<\sigma(2)<\dots <\sigma(p)$ and
$\sigma(p+1)<\dots <\sigma(n)$.
Notice that we always have $D=D\i1D\i2$.

With this definition, we know the coproduct for a basis of
$\calD$, the coproduct of a general term of $\calD$
is obtained by linearity:
$\Delta (\lambda D+\lambda' D') =
\lambda (\Delta D) + \lambda' (\Delta D')$.

The most important property of the coproduct is
its \emph{coassociativity}. We saw that the
coproduct of an element $D$ gives the ways to split
$D$ into two elements $D\i1$ and $D\i2$. 
Now assume that we want to split $D$ into
three elements. We can achieve this either by splitting
$D\i1$ or by splitting $D\i2$. Coassociativity
means that the result does not depend on this choice.
This is expressed more formally by
$(\Id\otimes\Delta)\Delta=(\Delta\otimes\Id)\Delta$.
For example the reader can check that
\begin{eqnarray*}
(\Id\otimes\Delta)\Delta \un &=& \un\otimes \un \otimes \un
= (\Delta\otimes\Id)\Delta \un,\\
(\Id\otimes\Delta)\Delta \partial &=& 
\partial\otimes \un \otimes \un + \un\otimes\partial\otimes \un +
\un\otimes \un \otimes \partial
= (\Delta\otimes\Id)\Delta \partial.
\end{eqnarray*}
The coproduct $\Delta$ is coassociative for all elements of $\calD$
\cite{Eisenbud}. 
It can also be shown that the coproduct satisfies
$\Delta D=\sum D\i1\otimes D\i2=\sum (-1)^{|D\i1||D\i2|} D\i2\otimes D\i1$
(this property is called graded cocommutativity).

We can define recursively the splitting of $D$ into
$n$ parts by $\Delta^{(0)}D=\un$, $\Delta^{(1)}D=D$,
$\Delta^{(2)}D=\Delta D$
and $\Delta^{(n)}D=(\Delta\otimes \Id^{n-2})\Delta^{(n-1)} D$
for $n>2$.
The result of the action of $\Delta^{(n)}$ on $D$ is denoted by
\begin{eqnarray}
\Delta^{(n)} D &=& \sum D\i1 \otimes\dots\otimes D\i{n}.
\label{Deltan}
\end{eqnarray}

To make a Hopf algebra, we need also a counit and
an antipode, but we shall not use these concepts in the
present paper. 

\subsection{The derivative of a product}
To show immediately the power of the Hopf algebraic concepts,
we prove the following formula for the derivative of
a product of two functions. If $D\in\calD$ is a product
of functional derivatives and 
$u$ and $v$ are functions of Dirac fields and sources
we have
\begin{eqnarray}
D (uv) &=& \sum (-1)^{|D\i2||u|} (D\i1 u)(D\i2 v).
\label{d(uv)}
\end{eqnarray}
In this equation, $|u|$ is the parity of the function
$u$. The parity of a function is defined as follows.
We first define the degree of a function: for a Dirac
field or a fermion source we have
$\degres(\psi)=\degres(\barpsi)=\degres(\eta)=\degres(\bareta)=1$.
The degree of a 
product of fields and sources is the sum of the
degrees of the fields and sources:
$\degres(uv)=\degres(u)+\degres(v)$, and the parity
of a function of fields and sources is equal to
the 0 or 1 when its degree is even or odd.
Notice that, if $\degres(D)\le\degres(u)$
we have $|Du|=|u|+|D|$ modulo 2 because
$\degres(Du)=\degres(u)-\degres(D)$.
The proof of (\ref{d(uv)})
is recursive. Equation (\ref{d(uv)})
is true for $D=\un$ because $\un(uv)=uv$ and for 
$D=\partial$ because of Leibniz' rule (\ref{Leibniz}).
If this is true for all elements of degree
up to $n$, take $D$ an element of
degree $n$ and define $D'=\partial D$.
On the one hand
\begin{eqnarray}
D' (uv) &=& \partial \big(D (uv))
=\sum (-1)^{|D\i2||u|} \partial
             \big((D\i1 u)(D\i2 v)\big) 
\nonumber\\
&=&
\sum (-1)^{|D\i2||u|} 
          (\partial D\i1 u)(D\i2 v)
\nonumber\\&&
          + (-1)^{|D\i2||u|+|D\i1|+|u|}
            (D\i1 u)(\partial D\i2 v)\big).
\label{intermed}
\end{eqnarray}
To obtain the last line, we used Leibniz' rule
and the fact that $|D\i1 u|=|u|+|D\i1|$ modulo 2.
On the other hand, by equation (\ref{DeltaDD'})
\begin{eqnarray*}
\Delta (\partial D) &=&
\sum (\partial  D\i1)\otimes D\i2
+ (-1)^{|D\i1|} D\i1\otimes (\partial D\i2).
\end{eqnarray*}
So that, if equation (\ref{d(uv)}) is true,
\begin{eqnarray*}
D' (uv) &=&
\sum (-1)^{|D\i2||u|}
          (\partial D\i1 u)(D\i2 v)
\\&&\hspace*{8mm}
          + (-1)^{|D\i1| + (1+|D\i2|)|u|}
            (D\i1 u) (\partial D\i2 v).
\end{eqnarray*}
But this is indeed equal to (\ref{intermed}), so
equation (\ref{d(uv)}) is satisfied for $D'$.
Since the elements $\partial D$ generate
$\calD_{n+1}$, equation (\ref{d(uv)}) is
true for $\calD$.
 
More generally
\begin{eqnarray}
D (u_1\dots u_n) &=& \sum (-1)^{\sum_{k=2}^n\sum_{l=1}^{k-1}
  |D\i{k}||u_l|} (D\i1 u_1)\dots (D\i{n} u_n).
\label{d(u1un)}
\end{eqnarray}
The recursive proof is left to the reader.

\subsection{Elimination of closed shells}
As a second application, we calculate $\tr(\hat\rho N^0)$ when
the system is composed of closed shells and
open shells. A closed shell is an electron
state $i_k$ which is occupied in all states
$|K\rangle$.  The open shells are the electron states
which are present in some but not all states $|K\rangle$.
Thus, the closed and open shells have no electron
state in common.
We rewrite equation (\ref{alternativex})
as $\tr(\hat\rho N^0)=\ee^d (uv)$ where
$d=\sum_n \partial^2/\partial \alpha_n \partial \bara_n$,
$u=\bara_{m_1}\alpha_{m_1}\dots\bara_{m_C}\alpha_{m_C}$
describes the closed shells containing $C$ electrons
and
$v= \sum_{KL} \rho_{LK}\bara_{j_1} \alpha_{i_1}\dots \bara_{j_N} \alpha_{i_N}$
describes the open shells.
Notice that in $u$ the index of each $\bara$ is the
same as the index of the following $\alpha$. This is
because the electron states are ordered so that
the closed shell have an index smaller than the open shells,
and the closed shells are occupied in all
$|K\rangle$ and $|L\rangle$.
To calculate $\tr(\hat\rho N^0)$ we first compute
$d(uv)$. According to equation (\ref{d(uv)})
\begin{eqnarray}
d(uv) &=& \sum (-1)^{|d\i2||u|}
(d\i1 u)(d\i2 v)= \sum (d\i1 u)(d\i2 v),
\label{dd(uv)}
\end{eqnarray}
because $|u|=2C=0$ modulo 2.
Now
\begin{eqnarray*}
\Delta d &=&
\sum_n \frac{\partial^2}{\partial \alpha_n
\partial \bara_n} \otimes \un +
\frac {\partial}{\partial \alpha_n}\otimes
\frac {\partial}{\partial \bara_n}
-
\frac {\partial}{\partial \bara_n}\otimes
\frac {\partial}{\partial \alpha_n} +
\un\otimes
\frac{\partial^2}{\partial \alpha_n
\partial \bara_n}.
\end{eqnarray*}
The terms $\partial u/\partial\alpha_n \partial v/\partial\bara_n$
and $\partial u/\partial\bara_n \partial v/\partial\alpha_n$
in equation (\ref{dd(uv)})
are zero because the closed and open shells have no
state in common. Therefore
$d(uv)=(du)v + u(dv)$. Moreover,
\begin{eqnarray*}
du &=& \sum_{k=1}^C
\frac{u}{\bara_{m_{k}}\alpha_{m_{k}}},
\end{eqnarray*}
is a sum of closed shells, so we can apply the
same argument again to show that
\begin{eqnarray*}
d^k(uv) &=& \sum_{l=0}^k {k\choose{l}}
(d^l u)(d^{k-l} v).
\end{eqnarray*}
Therefore
\begin{eqnarray*}
\tr(\hat\rho N^0) &=& \ee^d (uv)
=\sum_{k=0}^\infty \frac{1}{k!} d^k(uv)
=\sum_{k=0}^\infty  \sum_{l=0}^k \frac{1}{l!(k-l)!} (d^l
u)(d^{k-l} v),
\\&=&
\sum_{l=0}^\infty  \frac{1}{l!} d^lu \sum_{m=0}^\infty
\frac{1}{m!} d^mv
=(\ee^d u) (\ee^d v)
=\prod_{i=1}^C (1+\bara_{m_i}\alpha_{m_i}) (\ee^d v).
\end{eqnarray*}
In other words, the closed shell factorize in
$\tr(\hat\rho N^0)$. This result will be important to
restrict the size of the problem.
 
Notice that, in the proof, we used only the fact that
the closed and open shells have no electron state
in common. So the same reasoning shows that, if
the system is composed of two independent subsystems,
then $N^0_{KL}$ is the product of the $N^0_{KL}$
of both systems. More precisely, if all states
can be written as $|K\rangle = |K_1\rangle\wedge |K_2\rangle$,
where $\wedge$ antisymmetrizes the electron states
of $|K_1\rangle$ and $|K_2\rangle$,
where $|K_1\rangle$ has the same number of
electron states for all $|K\rangle$ and where
no $|K_1\rangle$ and $|K'_2\rangle$ have
any electron state in common for any
$|K\rangle$ and $|K'\rangle$, then
$N^0_{KL}=N^0_{K_1L_1}N^0_{K_2L_2}$.

\section{Calculation of $W^0_\rho$}
\label{CalcW0sect}
It will be very useful to define 
$W^0_\rho=\log(Z^0_\rho)$.
If the system has $N+C$ electrons with
$C$ electrons in closed shells, $Z^0_\rho$ 
can be written
\begin{eqnarray}
Z^0_\rho&=& \exp[-i\int \bareta(x) G^0_0(x,y)\eta(y)\dd x\dd y]
\prod_{i=1}^C (1+\bara_{m_i}\alpha_{m_i}) \rho(\bara,\alpha),
\label{Z0rho}
\end{eqnarray}
with
\begin{eqnarray}
\rho(\bara,\alpha) &=& 
\sum_{k=0}^N \rho_k(\bara,\alpha),
\label{rho(baraalpha)}
\end{eqnarray}
where $\rho_k$ contains products of $k$ $\bara$ and
$k$ $\alpha$. More explicitly
\begin{eqnarray}
\rho_N(\bara,\alpha)
&=& \sum \rho_{j_N\dots j_1,i_N\dots i_1}
\bara_{j_1} \alpha_{i_1}
\dots
\bara_{j_N} \alpha_{i_N},
\label{rhoN}
\\
\rho_k(\bara,\alpha) &=&
\frac{1}{(N-k)!}\Big(\sum_n \frac{\partial^2}{\partial \alpha_n
\partial \bara_n}\Big)^{N-k} \rho_N(\bara,\alpha).
\label{rhok}
\end{eqnarray}
In particular,
\begin{eqnarray}
\rho_0(\bara,\alpha)
&=& \tr(\hat\rho),\nonumber\\
\rho_1(\bara,\alpha)
&=&
\sum \rho_{j_N\dots j_1,i_N\dots i_1}
\Big( \sum_{l=1}^N \bara_{j_l} \alpha_{i_l}
  \prod_{p\not= l} \delta_{j_p,i_p}
\nonumber\\&&\hspace*{-15mm}
+
\sum_{l=1}^{N-1} \sum_{m=l+1}^N (-1)^{l+m}
\bara_{j_l} \alpha_{i_m}
  \prod_{p<l} \delta_{j_p,i_p}
  \prod_{l<p\le m} \delta_{j_p,i_{p-1}}
  \prod_{p>m} \delta_{j_p,i_p}
\nonumber\\&&\hspace*{-15mm}
+
\sum_{l=2}^{N} \sum_{m=1}^{l-1} (-1)^{l+m}
\bara_{j_l} \alpha_{i_m}
  \prod_{p<m} \delta_{j_p,i_p}
  \prod_{m\le p< l} \delta_{j_p,i_{p+1}}
  \prod_{p>l} \delta_{j_p,i_p}\Big),
\label{rho1}
\end{eqnarray}
will be useful.
It is important to isolate $\rho_1(\bara,\alpha)$,
which depends linearly on $\bara$ and $\alpha$,
because it will become a part of the free propagator.

The closed shells are dealt with easily:
\begin{eqnarray*}
\log\Big(\prod_{i=1}^C (1+\bara_{m_i}\alpha_{m_i})\Big)
&=& 
\sum_{i=1}^C \log(1+\bara_{m_i}\alpha_{m_i})
\\&=&
\sum_{i=1}^C\sum_{n=1}^\infty
\frac{(-1)^{n+1}}{n} (\bara_{m_i}\alpha_{m_i})^n.
\end{eqnarray*}
However, $\bara_{m_i}$ and $\alpha_{m_i}$ are fermionic
variables, thus
$(\bara_{m_i}\alpha_{m_i})^2=
\bara_{m_i}\alpha_{m_i}\bara_{m_i}\alpha_{m_i}
=-\bara_{m_i}\bara_{m_i}\alpha_{m_i}\alpha_{m_i}=0$
because, as fermionic variables,
$\bara_{m_i}^2=\alpha_{m_i}^2=0$.
Consequently, only the term $n=1$ remains in the sum
and
\begin{eqnarray*}
\log\Big(\prod_{i=1}^C (1+\bara_{m_i}\alpha_{m_i})\Big)
&=& 
\sum_{i=1}^C \bara_{m_i}\alpha_{m_i}.
\end{eqnarray*}
This result is important because it justifies the
fact that the propagator of the Green function
in many-body theory is obtained by summing the
contribution of all occupied shells. We see now that
this procedure is justified when the vacuum
$|\Phi_0\rangle$  can be
written as a full shell. In all other cases, this
procedure must be modified. The modification
comes from the term $\rho(\bara,\alpha)$ that we write
\begin{eqnarray*}
\rho(\bara,\alpha) &=& \tr(\hat\rho) + 
\sum_{k=1}^N \rho_k(\bara,\alpha)=
\tr(\hat\rho)\big(1+\sum_{k=1}^N 
\frac{\rho_k(\bara,\alpha)}{\tr(\hat\rho)}
\big).
\end{eqnarray*}
The usual convention is to impose $\tr(\hat\rho)=1$,
but we want to relax this constraint for later convenience.
Thus
\begin{eqnarray*}
\log(\rho(\bara,\alpha)) 
&=&
\log(\tr(\hat\rho))
+\log\big(1+\sum_{k=1}^N 
\frac{\rho_k(\bara,\alpha)}{\tr(\hat\rho)}\big),
\\&=&
\log(\tr(\hat\rho))+
\frac{\rho_1(\bara,\alpha)}{\tr(\hat\rho)}
+\rho^c(\bara,\alpha),
\end{eqnarray*}
where $\rho^c(\bara,\alpha)$ is defined by the last equation.
We can write $\rho^c(\bara,\alpha)$ as
\begin{eqnarray}
\rho^c(\bara,\alpha) &=& 
\sum_{n=2}^\infty 
\rho^c_n(\bara,\alpha),
\label{rhoc(baraalpha)}
\end{eqnarray}
where $\rho^c_n$ is the sum of the terms of $\rho^c$
which have degree $n$ in $\bara$ and degree $n$
in $\alpha$.
Notice that the sum over $n$ is finite. 
For instance, if the states
$|K\rangle$ are built by choosing $N$ electron orbitals
among $M$ (for instance, for Cr$^{3+}$, we have
three d electrons so that $N=3$ and $M=10$).
Therefore, $\rho_n(\bara,\alpha)^{1+M/n}=0$ because
in each term of $\rho_n(\bara,\alpha)^{1+M/n}$ 
at least one $\alpha_{i_p}$ is found twice and
$\alpha_{i_p}^2=0$. Therefore, $\rho^c_n(\bara,\alpha)=0$
for $n>M$.

If we gather all these results we obtain that
\begin{eqnarray}
W_\rho^0 &=& \log(Z_\rho^0)
=
-i\int \bareta(x) G^0_0(x,y)\eta(y)\dd x\dd y
+\sum_{i=1}^C \bara_{m_i}\alpha_{m_i}
\nonumber\\&&\hspace*{16mm}
+\log(\tr(\hat\rho))+
\frac{\rho_1(\bara,\alpha)}{\tr(\hat\rho)}
+\rho^c(\bara,\alpha),
\label{defWrho0}
\end{eqnarray}
where we recall that
$\bara_n=\int (\bareta_+(x)-\bareta_-(x))u_n(x)\dd x$
and $\alpha_n=\int \baru_n(x)(\eta_+(x)-\eta_-(x))\dd x$.
The term containing $G^0_0(x,y)$ is linear in
$\bareta$ and $\eta$. Thus, we shall include the
other linear terms by defining
\begin{eqnarray*}
G^0_\rho(x,y) &=& G^0_0(x,y) 
+i\big(\sum_{i=1}^C u_{m_i}(x) \bar{u}_{m_i}(y)
+\frac{\rho_1(x,y)}{\tr(\hat\rho)}\big)
\left( \begin{array}{cc}
    1 & -1 \\
   -1 & 1 \end{array}\right),
\end{eqnarray*}
with $\rho_1(x,y)$ defined so that
$\rho_1(\bara,\alpha)=\int \bareta_d(x) \rho_1(x,y) \eta_d(y)
\dd x\dd y$, in other words, $\rho_1(x,y)$
is obtained by replacing all $\bara_{i_k}$ by $u_{i_k}(x)$
and all $\alpha_{j_k}$ by $\bar{u}_{j_k}(y)$
in equation (\ref{rho1}).
It is at this stage that, when the system has only
closed shells, the effect of the closed shells is
entirely taken into account by adding the
occupied orbitals to the free Green function.
This procedure, which is universally used in the
quantum many-body approach, is usually deduced
fromt the particle-hole transformation.
This transformation is itself justified by 
showing that the Hamiltonian without
interaction $H_0$ is left invariant
(up to a pure number) \cite{Gross}.
However, this justification falls short of
being a proof that this procedure is
valid at all orders of the interacting
theory. From the previous discussion,
we see that the procedure is correct at
all orders when the noninteracting system
can be described by a single Slater determinant
(i.e. a closed shell).
However, the most interesting
phenomenon occurs when open shells are present.
We rewrite
\begin{eqnarray}
W_\rho^0 &=& 
-i\int \bareta(x) G^0_\rho(x,y)\eta(y)\dd x\dd y
+\log(\tr(\hat\rho))+
\rho^c(\bara,\alpha).
\label{Wrho0fin}
\end{eqnarray}
This is the final result of the section.

\section{The Green function hierarchy}
In this section, the Green function hierarchy
is established in the presence of open shells.

\subsection{Definition of Green functions}
According to the discussion of \ref{transampsect},
the expectation value of the Heisenberg field
$\psi_H(x)$ is given by
\begin{eqnarray}
\langle \psi_H(x)\rangle_\rho
&=& -i\frac{\delta Z_\rho}{\delta \bareta_+(x)}|_{\bareta=\eta=0}.
\label{psiH1}
\end{eqnarray}
The density matrix is normalised by $\tr(\hat\rho)=1$,
so that $Z_\rho|_{\bareta=\eta=0}=1$. Therefore,
we can also define
\begin{eqnarray}
\langle \psi_H(x)\rangle_\rho
&=& \Big(\frac{1}{Z_\rho}
\frac{-i\delta Z_\rho}{\delta \bareta_+(x)}\Big)|_{\bareta=\eta=0}.
\label{psiH2}
\end{eqnarray}
Although these definitions are equivalent, equation (\ref{psiH2})
has some advantages over equation (\ref{psiH1}):
(i) If we multiply $\rho$ by $\lambda$, equation (\ref{psiH2})
is not changed because the factor
$\lambda$ is cancelled between the numerator and the 
denominator. Thus, it is possible to relax the constraint
$\tr(\hat\rho)=1$ and we are enabled to consider unconstrained
density matrix. In particular, we can use
$\hat\rho=\exp[-\beta H]$ for equilibrium quantum field
theory. (ii) If equations (\ref{psiH1}) and (\ref{psiH2}) 
are written as a sum of Feynman diagrams, 
equation (\ref{psiH1}) has vacuum diagrams which
are cancelled by the denominator of equation (\ref{psiH2}),
in other words, only equation (\ref{psiH2}) is a sum of connected
diagrams. (iii) When the density matrix $\hat\rho$ is
that of the vacuum (i.e. $\hat\rho=|0\rangle\langle 0|$),
equation (\ref{psiH2}) is the Gell-Mann and Low equation
\cite{GellMann} which is known to be correct.
(iv) Equation (\ref{psiH2}) has been used successfully since
the early days of nonequilibrium quantum field theory
\cite{Keldysh}.

It turns out that a complete set of equations cannot
be obtained by functional derivatives with respect
to $\eta_+$ and $\bareta_+$ alone. So we define the
following expectations values:
\begin{eqnarray*}
\langle \psi_+(x)\rangle_\rho &=&
\langle \psi_H(x)\rangle_\rho
= \Big(\frac{1}{Z_\rho}
\frac{-i\delta Z_\rho}{\delta \bareta_+(x)}\Big)|_{\bareta=\eta=0},
\\
\langle \barpsi_+(x)\rangle_\rho &=&
\langle \barpsi_H(x)\rangle_\rho
= \Big(\frac{1}{Z_\rho}
\frac{i\delta Z_\rho}{\delta \eta_+(x)}\Big)|_{\bareta=\eta=0},
\\
\langle \psi_-(x)\rangle_\rho &=&
 \Big(\frac{1}{Z_\rho}
\frac{i\delta Z_\rho}{\delta \bareta_-(x)}\Big)|_{\bareta=\eta=0},
\quad
\langle \barpsi_-(x)\rangle_\rho =
 \Big(\frac{1}{Z_\rho}
\frac{-i\delta Z_\rho}{\delta \eta_-(x)}\Big)|_{\bareta=\eta=0}.
\end{eqnarray*}

\subsection{Hierarchy of disconnected Green functions}
We rewrite equation(\ref{Zrho=DZ0rho}) as
$Z_\rho=\ee^{-iD}Z^0_\rho$ where
\begin{eqnarray}
D &=&
\int H^\inter(\frac{i\delta}{\delta\bareta_+(x)},
               \frac{-i\delta}{\delta\eta_+(x)})
- H^\inter(\frac{-i\delta}{\delta\bareta_-(x)},
               \frac{i\delta}{\delta\eta_-(x)}) \dd x.
\label{defD}
\end{eqnarray}
The operator $D$ contains products of 2 or 4 functional
derivatives, thus $D$ is even and $D$ commutes
with the elements of $\calD$. Thus, if
$\beta=\eta_\pm(x)$ or $\beta=\bareta_\pm(x)$
\begin{eqnarray*}
\frac{\delta Z_\rho}{\delta\beta} &=&
\ee^{-iD} \frac{\delta Z^0_\rho}{\delta\beta}.
\end{eqnarray*}
We use the fact that $Z^0_\rho=\ee^{W^0_\rho}$
with $|W^0_\rho|=0$ to get
\begin{eqnarray}
\frac{\delta Z_\rho}{\delta\beta} &=&
\ee^{-iD} \Big(\frac{\delta W^0_\rho}{\delta\beta}\ee^{W^0_\rho}\Big)
=\ee^{-iD} \Big(\frac{\delta W^0_\rho}{\delta\beta}Z^0_\rho\Big)
\nonumber\\&=&
\sum_{n=0}^\infty \frac{(-i)^n}{n!} D^n \big(
\frac{\delta W^0_\rho}{\delta\beta}Z^0_\rho\big).
\label{nonreduit}
\end{eqnarray}
The action of the operator $D^n$ is expanded with
equation (\ref{d(uv)}), using $|\delta W^0_\rho/\delta\beta|=1$:
\begin{eqnarray*}
\frac{\delta Z_\rho}{\delta\beta} &=&
\sum_{n=0}^\infty \frac{(-i)^n}{n!} 
\sum (-1)^{|D\i2^n|} \big(D\i1^n 
\frac{\delta W^0_\rho}{\delta\beta}\big)
\big(D\i2^n Z^0_\rho\big).
\end{eqnarray*}
We transform this infinite sum into a finite
sum by using reduced coproducts.
The reduced coproduct with respect to $D$ is denoted by $\Delta'D$.
It is defined as follows, the reduced coproduct with respect to $D$
of $D$ itself is defined by 
$\Delta'D = \Delta D - 1\otimes D - D\otimes 1$.
The Sweedler notation for it is
$\Delta' D = \sum D\i{1'}\otimes D\i{2'}$. The
reduced coproduct of $D^n$ is defined recursively
by
\begin{eqnarray}
\Delta'(D^{n+1}) &=&
\sum (-1)^{|D\i{1'}||D\i{2'}^n|} 
D\i{1'}^nD\i{1'}\otimes D\i{2'}^nD\i{2'}.
\label{Delta'def}
\end{eqnarray}
This is extended to $n=0$ by $\Delta'(D^0)=1\otimes 1$.
An equivalent definition is that $\Delta'(D^{n})$
is the sum of all terms of $\Delta (D^n)$ which do 
not contain any $D$.
The relation between $\Delta (D^n)$ and $\Delta'(D^n)$
is given by 
\begin{eqnarray}
\Delta (D^n) &=&
\sum_{k=0}^n\sum_{l=0}^{n-k}
\frac{n!}{k!l!(n-k-l)!} D\i{1'}^{n-k-l} D^k
                       \otimes D\i{2'}^{n-k-l} D^l.
\label{DeltaDelta'}
\end{eqnarray}
This can be shown by a recursive proof. The definition of
$\Delta'D$ gives us
\begin{eqnarray}
\Delta D &=& D\otimes 1 + 1\otimes D + \sum D\i{1'}\otimes D\i{2'},
\label{DeltaDelta'D}
\end{eqnarray}
so equation (\ref{DeltaDelta'}) is true for $n=1$.
Assume that it is true for all $D^k$ for all $k$ up to $n$.
From equations (\ref{DeltaDD'}), (\ref{DeltaDelta'}) and
(\ref{DeltaDelta'D}) we obtain (using $|D|=0$),
\begin{eqnarray*}
\Delta(D^nD) &=& \sum_{k+l+m=n} \frac{n!}{k!l!m!}
\Big(D\i{1'}^{m} D^{k+1} \otimes D\i{2'}^{m} D^l
\\&&
+D\i{1'}^{m} D^{k} \otimes D\i{2'}^{m} D^{l+1}
\\&&
+(-1)^{|D\i{1'}||D\i{2'}^{m}|} D\i{1'}^{m} D\i{1'} D^k 
                \otimes D\i{2'}^{m} D\i{2'}D^l\Big).
\end{eqnarray*}
Using the recursive definition (\ref{Delta'def}) we get
\begin{eqnarray*}
\Delta(D^nD) &=& \sum_{k+l+m=n} \frac{n!}{k!l!m!}
\Big(D\i{1'}^{m} D^{k+1} \otimes D\i{2'}^{m} D^l
\\&&
+D\i{1'}^{m} D^{k} \otimes D\i{2'}^{m} D^{l+1}
+D\i{1'}^{m+1} D^k \otimes D\i{2'}^{m+1} D^l\Big).
\end{eqnarray*}
This can be rewritten
\begin{eqnarray*}
\Delta(D^nD) &=& \sum_{k+l+m=n+1}\Big( 
\frac{n!}{(k-1)!l!m!}+ \frac{n!}{k!(l-1)!m!}
+ \frac{n!}{k!l!(m-1)!}\Big)
\\&&\hspace*{15mm}
D\i{1'}^{m} D^k \otimes D\i{2'}^{m} D^l.
\end{eqnarray*}
The first three integers can be summed to
\begin{eqnarray*}
\Delta(D^nD) &=& \sum_{k+l+m=n+1}
\frac{n!(k+l+m)}{k!l!m!} D\i{1'}^{m} D^k \otimes D\i{2'}^{m} D^l,
\end{eqnarray*}
and equation (\ref{DeltaDelta'}) is proved for
$D^{n+1}$.

By summing equation (\ref{DeltaDelta'}) over $n$
we obtain the important identity
\begin{eqnarray}
\Delta \ee^D &=& 
\sum_{n=0}^\infty \frac{1}{n!}  D\i{1'}^{n} \ee^D \otimes D\i{2'}^{n} \ee^D
= 
(\Delta' \ee^D) (\ee^D\otimes \ee^D).
\label{DeltaeD}
\end{eqnarray}
Note that this identity is true for any graded commutative Hopf algebra
and any $D$ of degree $>0$.

Using identity (\ref{DeltaeD}), the equation (\ref{nonreduit}) for 
$\delta Z_\rho/\delta\beta$ becomes
\begin{eqnarray*}
\frac{\delta Z_\rho}{\delta\beta} &=&
\sum_{n=0}^\infty \frac{(-i)^n}{n!}
\sum (-1)^{|D\i{2'}^{n}|} \big(D\i{1'}^{n} 
\ee^{-iD} \frac{\delta W^0_\rho}{\delta\beta}\big)
\big(D\i{2'}^{n} \ee^{-iD} Z^0_\rho\big),
\\&=&
\sum_{n=0}^\infty \frac{(-i)^n}{n!}
\sum (-1)^{|D\i{2'}^{n}|} \big(D\i{1'}^{n} 
\frac{\delta W^1_\rho}{\delta\beta}\big)
\big(D\i{2'}^{n} Z_\rho\big),
\end{eqnarray*}
where $W^1_\rho=\ee^{-iD} W^0_\rho$
adds the electron-electron interactions to
the cumulant  $W^0_\rho$ of the moment generating 
function $Z^0_\rho$.
Since the cumulant $W^0_\rho$ is a finite polynomial
in $\bara$ and $\alpha$, 
the interacting cumulant $W^1_\rho$ is also a
finite polynomial in $\bara$ and $\alpha$.
Now each  $D\i{1'}^{m}$ (for $m\not=0$) contains
at least $m$  functional derivatives with
respect to $\bareta_\pm$ or $\eta_\pm$
(this is why the reduced coproduct was defined),
thus $D\i{1'}^{m}$ is zero for $m$ large enough.
In fact, $m=2M-1$ is a possible bound and
we obtain our final formula,
isolating the contribution of $n=0$
\begin{eqnarray}
\frac{\delta Z_\rho}{\delta\beta} &=&
\frac{\delta W^1_\rho}{\delta\beta} Z_\rho+
\sum_{n=1}^{2M-1} \frac{(-i)^n}{n!}
\sum (-1)^{|D\i{2'}^{n}|} \big(D\i{1'}^{n} 
\frac{\delta W^1_\rho}{\delta\beta}\big)
\big(D\i{2'}^{n} Z_\rho\big).
\nonumber\\
\label{reduit}
\end{eqnarray}
We have transformed the infinite 
sum (\ref{nonreduit}) into the finite
sum (\ref{reduit}).
To be complete, we still have to 
replace the disconnected Green functions defined 
by functional derivatives with respect to $Z_\rho$
by connected Green functions defined by
functional derivatives with respect
to $W_\rho=\log Z_\rho$.

\subsection{Calculation of $W^1_\rho$}
\label{W0W1}
Apparently, $W^1_\rho=\ee^{-iD} W^0_\rho$ 
includes some interaction in $W^0_\rho$, but
in the interaction Hamiltonian $H^\inter$ that we consider,
we have $W^1_\rho=W^0_\rho$. 
Indeed, these contain integrals over
$d=\delta^2/\delta{\eta_\pm}(x)\delta{\bareta_\pm}(x)$.
The action of $d$ on the term containing the
Green function $G^0_\rho(x,y)$ is irrelevant because it
gives a term independent of $\alpha$ and $\bara$.
For the action on $\rho^c(\bara,\alpha)$ we have
\begin{eqnarray*}
\frac{\delta\rho^c}{\delta\bareta_\pm(x)} 
&=& \sum_n \frac{\partial\rho^c}{\partial\alpha_n} 
  \frac{\delta\alpha_n}{\delta\bareta_\pm(x)}
= \pm \sum_n \frac{\partial\rho^c}{\partial\alpha_n} u_n(x),
\end{eqnarray*}
\begin{eqnarray*}
\frac{\delta^2\rho^c}{\delta{\eta_\pm}(x)\delta{\bareta_\pm}(x)}
&=& \sum_{mn} \frac{\partial^2\rho^c}{\partial\bara_m\partial\alpha_n} \baru_m(x) u_n(x).
\end{eqnarray*}
Remark that the right hand side of the last equation does not depend on the
sign $\pm$ of the source.
The diffential operator $D$ can be written as $D=D_+-D_-$,
where $D_+$ and $D_-$ are the same operators, but the
first one involves derivatives with respect to the $+$ sources
and the second one with respect to the $-$ sources.
According to our remark, $D_+\rho^c=D_-\rho^c$.
Thus, $D\rho^c=0$ and $W^1_\rho=W^0_\rho$.

\subsection{Hierarchy of connected Green functions}
In formula (\ref{reduit}), the differential
operator $D\i{2'}^{n}$ acts on 
$Z_\rho=\ee^{W_\rho}=\sum_{n=0}^\infty W_\rho^n/n!$.
Thus, we must determine the action of a differential
operator on $W_\rho^n$. Notice that $|Z_\rho|=0$,
thus $|W_\rho|=0$.

So we take an even element $u$ (even means that $|u|=0$)
and a differential operator $d$ such that $\degres(d)>0$
and we want to calculate $d u^n$.
We shall use now the standard reduced coproduct
$\Deltau$ defined, for any element $d\in\calD$ by
$\Deltau d =\Delta d - d\otimes 1 - 1\otimes d$,
and we write $\Deltau d= \sum d\iu1 \otimes d\iu2$.
This reduced coproduct is coassociative.
The basic identity that we need is
\begin{eqnarray}
d(u^n) &=& \sum_{k=1}^n 
{n\choose{k}} u^{n-k} \sum d\iu{1}u\dots d\iu{k}u,
\label{dun}
\end{eqnarray}
where $d\iu{1}u=du$ if $k=1$
and $\Deltau^{(k)} d = \sum d\iu{1}\otimes\dots\otimes d\iu{k}$
is defined recursively from $\Deltau$ as in equation (\ref{Deltan}).
For example, using equation (\ref{d(uv)}) and $|u|=0$
\begin{eqnarray*}
d(u^2) &=& \sum (d\i1 u)(d\i2 u)
= (du) u + u (du) + \sum (d\iu{1}u)(d\iu{2}u),
\\&=& 2 u\, du + \sum (d\iu{1}u)(d\iu{2}u),
\end{eqnarray*}
and equation (\ref{dun}) is valid for $n=2$.
The general case is proved recursively. Assume
that it is true up to $n$, then
\begin{eqnarray*}
d(u^{n+1}) &=& \sum (d\i1 u^n)(d\i2 u)
= u^n du + d(u^n) u + \sum (d\iu{1}u^n) (d\iu{2}u),
\\&=&
 u^n du + \sum_{k=1}^n
{n\choose{k}} u^{n-k+1} \sum d\iu{1}u\dots d\iu{k}u
\\&&
+ \sum_{k=1}^n
{n\choose{k}} u^{n-k} \sum (d\iu{1}\iu{1}u\dots d\iu{1}\iu{k}u)
(d\iu{2}u)
\\&=&
 u^n du + \sum_{k=1}^n
{n\choose{k}} u^{n-k+1} \sum d\iu{1}u\dots d\iu{k}u
\\&&
+ \sum_{k=1}^n
{n\choose{k}} u^{n-k} \sum d\iu{1}u\dots d\iu{{k+1}}u
\\&=&
 u^n du + \sum_{k=1}^n
{n\choose{k}} u^{n-k+1} \sum d\iu{1}u\dots d\iu{k}u
\\&&
+ \sum_{k=2}^{n+1}
{n\choose{k-1}} u^{n-k+1} \sum d\iu{1}u\dots d\iu{k}u
\\ &=& \sum_{k=1}^{n+1}
{n+1\choose{k}} u^{n+1-k} \sum d\iu{1}u\dots d\iu{k}u.
\end{eqnarray*}
We used the coassociativity of the reduced coproduct.
From equation (\ref{dun}) we can calculate
\begin{eqnarray}
d(\ee^u) &=& \sum_{n=0}^\infty \frac{1}{n!} d(u^n)
=
\sum_{n=1}^\infty  \frac{1}{n!}
\sum_{k=1}^n 
{n\choose{k}} u^{n-k} \sum d\iu{1}u\dots d\iu{k}u,
\nonumber\\&=&
\sum_{m=0}^\infty  \frac{u^m}{m!}
\sum_{k=1}^\infty 
\frac{1}{k!} \sum d\iu{1}u\dots d\iu{k}u,
\nonumber\\&=&
\ee^u
\sum_{k=1}^\infty 
\frac{1}{k!} \sum d\iu{1}u\dots d\iu{k}u.
\label{deu}
\end{eqnarray}
The sum over $k$ is not infinite because
${\Deltau}^{(k)} d=0$ if $k>\degres(d)$ and the
sum stops at $k=\degres(d)$.
More generally, for an analytic function $f(z)$,
\begin{eqnarray*}
d\big(f(u)\big) &=& 
\sum_{k=1}^\infty 
\frac{f^{(k)}(u)}{k!} \sum d\iu{1}u\dots d\iu{k}u,
\end{eqnarray*}
where $f^{(k)}(u)$ is the $k$-th derivative
of $f$ at $u$. The cocommutativity of the coproduct
ensures that the factor $1/k!$ disappears from the expanded
formulas.

If equation (\ref{deu}) is applied to $u=W_\rho$, we obtain a
relation between unconnected Green functions $(1/Z_\rho) d Z_\rho$
and connected Green functions $d W_\rho$. For instance,
if $d=\partial\partial'$, then
$\Deltau d = \partial\otimes\partial'
-\partial'\otimes\partial$ and
$(1/Z_\rho) d Z_\rho= d W_\rho + (1/2) (\partial W_\rho)(\partial' W_\rho)
- (1/2) (\partial' W_\rho) (\partial W_\rho)$.
At $\bareta_\pm=\eta_\pm=0$ we obtain
$ d Z_\rho= d W_\rho$.
Similarly, if $d=\overline{\partial}\, \overline{\partial}' \partial\partial'$,
where $\partial$ and $\partial'$ are derivative with respect to $\eta$
and $\overline{\partial}$ and $\overline{\partial}'$ with respect to $\bareta$,
we find at $\bareta_\pm=\eta_\pm=0$,
$ d Z_\rho= d W_\rho - 
(\overline{\partial}\partial  W_\rho)
(\overline{\partial}'\partial' W_\rho)
+
(\overline{\partial}\partial' W_\rho)
(\overline{\partial}'\partial W_\rho).
$

Equation (\ref{deu}) is now introduced into (\ref{reduit}),
where we use the fact that $0=|D^n|=|D\i{1'}^{n}|+|D\i{2'}^{n}|$,
so that $|D\i{2'}^{n}|=|D\i{1'}^{n}|$:
\begin{eqnarray*}
\frac{\delta Z_\rho}{\delta\beta} &=&
\sum_{n=0}^{2M-1} \frac{(-i)^n}{n!}
\sum (-1)^{|D\i{1'}^{n}|} \big(D\i{1'}^{n} 
\frac{\delta W^1_\rho}{\delta\beta}\big)
\\&&
\times Z_\rho
\sum_{k=1}^\infty 
\frac{1}{k!} \sum \big(D\i{2'}^{n}\iu{1} W_\rho\dots \big)
\big( D\i{2'}^{n}\iu{k}W_\rho\big).
\end{eqnarray*}
Using again the definition of $W_\rho$ in terms
of $Z_\rho$, we obtain an equation involving only
the connected Green functions:
\begin{eqnarray}
\frac{\delta W_\rho}{\delta\beta} &=&
\frac{1}{Z_\rho}\frac{\delta Z_\rho}{\delta\beta}
\nonumber\\&=& 
\sum_{n=0}^{2M-1} \frac{(-i)^n}{n!}\sum_{k=1}^\infty
\sum \frac{(-1)^{|D\i{1'}^{n}|}}{k!} 
\big(D\i{1'}^{n} 
\frac{\delta W^1_\rho}{\delta\beta}\big)
\nonumber\\&&\hspace*{20mm}
(D\i{2'}^{n}\iu{1} W_\rho)\dots (D\i{2'}^{n}\iu{k}W_\rho).
\label{reduitc}
\end{eqnarray}
This sum is finite because, for each $n$, the sum over $k$ stops at
$k=\degres(D\i{2'}^{n})$.

\section{Conclusion}

This paper had two purposes: (i) to
determine the hierarchy of Green functions
for degenerate systems, and more generally
for systems whose initial state cannot
be written as a Slater determinant; (ii)
to show the power of quantum groups
and Hopf algebras to solve problems
of quantum field theory.
A detailed application of the formulas
obtained in this paper can be found
in reference \cite{BrouderKB2}.

In this paper we dealt with a nonrelativistic
electronic system with Coulomb interaction.
A generalization to quantum electrodyamics is possible,
which would provide an alternative to the
new methods recently developed to 
carry out quantum electrodynamical calculations of 
many-electron systems
\cite{Lindgren0,Lindgren1,Lindgren3,Lebigot,Shabaev}.
Again, the present method has the advantage
of being self-consistent and of preserving the
symmetry of the system.

Moreover, a functional derivation of the energy
with respect to the density matrix provides
equations that enable us to unify the Green-function
formalism and the diagonalization method of
many-body theory. This will be presented in
a forthcoming publication.

\section{Acknowledgements}
The answers of Volker Bach are
gratefully acknowledged, as well as the
fruitful comments by Matteo Calandra, Philippe Sainctavit
and Francesco Mauri.
I thank Emanuela Petracci for the kind and numerous
explanations she gave me on the Hopf algebra of derivations.
I gratefully acknowledge very fruitful discussions
with Florent Hivert, Nicolas Thiery and the
members of the Laboratoire d'Informatique at the
Marne la Vall{\'e}e University.
Many thanks to 
Sandrine Brice-Profeta,
Delphine Cabaret,
Emilie Gaudry,
Isabelle Letard,
Mayeul d'Avezac de Castera,
Matteo Calandra, 
Vladimir Dmitrienko,
Michele Lazzeri,
Mikael Profeta
for their patient listening and their
clever questions during the lectures
I gave on the subject of this paper.

\section{Appendix : Calculation of the trace}
The calculation of 
 $\tr[\hat\rho N^0(\bareta_d,\eta_d)]$
is an essential ingredient of this work.
We rewrite the density matrix as
$\hat\rho=\sum_{KL} \rho_{LK} |L\rangle\langle K|$,
where $|K\rangle$ and $|L\rangle$
are Slater determinants defined by
$|K\rangle = b^\dagger_{i_N}\dots b^\dagger_{i_1} |0\rangle$
and
$|L\rangle = b^\dagger_{j_N}\dots b^\dagger_{j_1} |0\rangle$.
Here
$b^\dagger_{i_k}$ and $b^\dagger_{j_l}$
are creation operators of the one-electron orbitals
indexed by $i_k$ and $j_l$. The indices are
ordered ($i_1<\dots<i_N$, $j_1<\dots<j_N$).
The total number of
electrons in the system is $N$. Moreover,
 $|0\rangle$ is the true vacuum of the system
(i.e. containing no electron).
We must calculate
$\tr[\hat\rho N^0(\bareta_d,\eta_d)]=\sum_{KL} \rho_{LK} N^0_{KL}$
with
\begin{eqnarray}
N^0_{KL} &=&
\langle K | {:} \exp\big(i\int \bareta_d(x)\psi(x) +
\bar\psi(x)\eta_d(x)\dd x\big)
{:}|L\rangle.
\label{N0KL}
\end{eqnarray}
The fields are expanded over time-dependent eigenstates
of the one-body Hamiltonian
\begin{eqnarray*}
\psi(x) &=& \sum_n b_n u_n(x), \quad 
\barpsi(x) = \sum_n b^\dagger_n u^\dagger_n,
\end{eqnarray*}
where $u_n(x)$ are the time-dependent
solutions defined in section \ref{unsection}
and $n$ is the index of the electron orbital,
$b_n, b^\dagger_n$ are the annihilation and creation operators
  of an electron in orbital $n$ \cite{Fetter}.

We can rewrite $N^0_{KL}$ as
\begin{eqnarray*}
N^0_{KL} &=&\sum_{l=0}^\infty \frac{i^l}{l!}
\langle K | {:} \Big(\sum_n \int \bareta_d(x)u_n(x)\dd x\, b_n +
    b^\dagger_n \int \baru(x)\eta_d(x)\dd x\Big)^l
{:}|L\rangle\\
&=&\sum_{l=0}^\infty \frac{i^l}{l!}
\langle K | {:} \Big(\sum_n \bara_n b_n +
    b^\dagger_n \alpha_n\Big)^l {:}|L\rangle,
\end{eqnarray*}
where $\bara_n=\int \bareta_d(x)u_n(x)\dd x$
and $\alpha_n=\int \baru_n(x)\eta_d(x)\dd x$
are anticommuting variables.
To calculate $N^0_{KL}$, we first notice that
the anticommutativity of $b_n,b^\dagger_n,\alpha_n$
and $\bara_n$ for the normal product gives us
the commutation rules
${:}\bara_i b_i\bara_{j} b_j{:}={:}\bara_{j} b_j\bara_i
b_i{:}$,
${:}\bara_i b_i b^\dagger_j \alpha_j{:}={:} b^\dagger_j
\alpha_j
  \bara_i b_i{:}$ and
${:} b^\dagger_i \alpha_i b^\dagger_j \alpha_j{:}={:}
  b^\dagger_j \alpha_j b^\dagger_i \alpha_i{:}$.
Thus, we can expand the power with the binomial formula
\begin{eqnarray*}
N_{KL}^0 &=&
\sum_{l=0}^\infty \frac{i^l}{l!} \sum_{k=0}^l {l\choose{k}}
\\&&
\sum_{n_1\cdots n_l}
\langle K |
 b^\dagger_{n_1} \alpha_{n_1}\dots b^\dagger_{n_k}
\alpha_{n_k}
 \bara_{n_{k+1}} b_{n_{k+1}}\dots \bara_{n_l} b_{n_l}
  |L\rangle\\
&=&  \sum_{l=0}^\infty \frac{i^l}{l!}
\sum_{k=0}^l {l\choose{k}} (-1)^{k+l(l-1)/2}
\sum_{n_1\cdots n_l}
\alpha_{n_1}\dots \alpha_{n_k}
\bara_{n_{k+1}}\dots \bara_{n_l}
\\&&
\langle K |
 b^\dagger_{n_1} \dots b^\dagger_{n_k}
  b_{n_{k+1}}\dots  b_{n_l}
  |L\rangle.
\end{eqnarray*}
The transition between $|K\rangle$ and
$|L\rangle$
is zero if $l\not= 2k$ or if $l>2N$ because
$|K\rangle$ and $|L\rangle$ contain $N$ electrons.
Thus we obtain the finite sum
\begin{eqnarray}
N_{KL}^0 &=& \sum_{k=0}^N \frac{(-1)^k}{(k!)^2}
\sum_{n_1\cdots m_{k}}
\alpha_{n_1}\dots \alpha_{n_k}
\bara_{m_{1}}\dots \bara_{m_{k}}
\nonumber\\&&
\langle K |
 b^\dagger_{n_1} \dots b^\dagger_{n_k}
  b_{m_{1}}\dots  b_{m_{k}}
  |L\rangle.
\label{an1ank}
\end{eqnarray}

\subsection{Hopf calculation}
 
Hopf algebraic techniques will be used to
obtain an explicit expression for $N_{KL}^0$.
We first denote
\begin{eqnarray*}
A_{KL} &=&
\langle K |
 b^\dagger_{n_1} \dots b^\dagger_{n_k}
  b_{m_{1}}\dots  b_{m_{k}}
  |L\rangle,
\end{eqnarray*}
and we write $u=b_{i_1}\dots b_{i_N}$,
$v=b^\dagger_{j_N}\dots b^\dagger_{j_1}$,
$s=b^\dagger_{n_1}\dots b^\dagger_{n_k}$
and
$t=b_{m_1}\dots b_{m_k}$.
Thus $A_{KL}=\langle 0| u({:}st{:})v|0\rangle$
and we use the Hopf version of Wick's theorem
\cite{BrouderQG}
\begin{eqnarray*}
({:}st{:})v &=& \sum (-1)^{|v\i1|(|s\i2|+|t\i2|)}
               ({:}st{:}\i1|v\i1) {:}({:}st{:}\i2 v\i2){:},
\\&=& \sum (-1)^{|v\i1||s\i2|+|v\i1||t\i2|+|t\i1||s\i2|}
               ({:}s\i1 t\i1{:}|v\i1) {:}s\i2 t\i2 v\i2{:}.
\end{eqnarray*}
Therefore
\begin{eqnarray*}
A_{KL} &=& \big(u|({:}st{:})v\big),
\\&=&
\sum (-1)^{|v\i1||s\i2|+|v\i1||t\i2|+|t\i1||s\i2|}
               ({:}s\i1 t\i1{:}|v\i1) (u|{:}s\i2 t\i2
v\i2{:}).
\end{eqnarray*}
 
In general
\begin{eqnarray}
({:}c_1\dots c_m{:}|{:} d_1\dots d_n{:}) &=& \delta_{m,n}
  (-1)^{n(n-1)/2} \det(M),
\label{detM}
\end{eqnarray}
where $c_i$ and $d_j$ are creation or annihilation operators
and $M$ is the $n\times n$ matrix with elements
$M_{ij}=(c_i|d_j)$ \cite{Grosshans}.
The Laplace pairing $(c_i|d_j)$ is obtained from
$(b_i|b^\dagger_j)=\delta_{i,j}$, $(b_i|b_j)=0$,
$(b^\dagger_i|b_j)=0$ and $(b^\dagger_i|b^\dagger_j)=0$.
Because of the value of $(b_i|d_j)$,
$({:}c_1\dots c_n{:}|{:} d_1\dots d_n{:})$ is
zero if any $c_i$ is a creation operator
or any $d_j$ an annihilation operator (because one
row or one column of $M$ is zero).
Therefore, we need $s\i1=1$ and $t\i2=1$, so that
$s\i2=s$ and $t\i1=t$:
\begin{eqnarray}
A_{KL} &=&
\sum (-1)^{|v\i1||s|+|t||s|}
               (t|v\i1) (u|{:}s v\i2{:}),
\\&=&
\sum (-1)^{|v\i1||s|+|t||s|+|u\i2||s|}
               (t|v\i1) (u\i1|s) (u\i2|v\i2).
\label{AKL}
\end{eqnarray}
We rewrite $v=(-1)^{N(N-1)/2}b^\dagger_{j_1}\dots
b^\dagger_{j_N}$
so that
\begin{eqnarray*}
\Delta u &=&
\sum_{p=0}^N \sum_{\sigma} (-1)^\sigma
b_{i_{\sigma(1)}}\dots b_{i_{\sigma(p)}}
\otimes  b_{i_{\sigma(p+1)}}\dots b_{i_{\sigma(N)}},\\
\Delta v &=& (-1)^{N(N-1)/2}
\sum_{q=0}^N \sum_{\tau} (-1)^\tau
b^\dagger_{j_{\tau(1)}}\dots b^\dagger_{j_{\tau(q)}}
\otimes  b^\dagger_{j_{\tau(q+1)}}\dots
b^\dagger_{j_{\tau(N)}},
\end{eqnarray*}
where $\sigma$ runs over the $(p,N-p)$-shuffles
and $\tau$ over the $(q,N-q)$-shuffles.
A $(p,N-p)$-shuffle is a permutation $\sigma$ of
$(1,\dots,N)$ such that
$\sigma(1)<\sigma(2)<\dots <\sigma(p)$
and $\sigma(p+1)<\dots <\sigma(N)$.
If $p=0$ or $p=N$, $\sigma$ is the identity permutation.
Equation (\ref{detM}) applied to (\ref{AKL}),
gives us $p=k$ and $q=k$
so that $|v\i1|=|s|=|t|=k$, $|u\i2|=N-k$ and
\begin{eqnarray}
A_{KL} &=& \sum (-1)^{N(N-1)/2+(N-k)k+k(k-1)+(N-k)(N-k-1)/2}
\nonumber\\&&
\sum_{\sigma\tau} (-1)^{\sigma+\tau}
\det(\delta_{m_p,j_{\tau(q)}})
\det(\delta_{i_{\sigma(p)},n_q})
\det(\delta_{i_{\sigma(p)},j_{\tau(q)}}),
\label{AKL2}
\end{eqnarray}
where $p$ and $q$ run from $1$ to $k$ in
the first two matrices and
from $k+1$ to $N$ in the last one.
The determinant of a $n\times n$ matrix $a_{ij}$ is
$\det(a)=\sum_{\lambda} (-1)^\lambda \prod_{i=1}^n
a_{i\lambda(i)}
=\sum_{\lambda} (-1)^\lambda \prod_{i=1}^n a_{\lambda(i)i}$,
where $\lambda$ runs over the permutations of $n$ elements.
Therefore, to calculate the last determinant in equation (\ref{AKL2}), 
we must sum over all permutations of $\tau(k+1),\dots,\tau(N)$,
but the indices satisfy $i_1<\dots<i_N$ and
$j_1<\dots<j_N$. By definition of the $(k,N-k)$-shuffle,
we have $i_{\sigma(k+1)}<\dots<i_{\sigma(N)}$ and
$j_{\tau(k+1)}<\dots<j_{\tau(N)}$ so any permutation
of $\tau(k+1)$,\dots,$\tau(N)$ would break this ordering
(for example, $\delta_{j_1,i_2}\delta_{j_2,i_1}=0$
because $j_1<j_2$ and $i_1<i_2$).
Thus the only nonzero term of
$\det(\delta_{i_{\sigma(p)},j_{\tau(q)}})$
is $\delta_{i_{\sigma(k+1)},j_{\tau(k+1)}}\dots
\delta_{i_{\sigma(N)},j_{\tau(N)}}$.
This gives us the following expression for $A_{KL}$:
\begin{eqnarray*}
A_{KL} &=& \sum (-1)^{k(k-1)/2}
\sum_{\sigma\tau} (-1)^{\sigma+\tau}
\det(\delta_{m_p,j_{\tau(q)}})
\det(\delta_{i_{\sigma(p)},n_q})
\\&&\hspace*{20mm}
\prod_{p=k+1}^N
\delta_{i_{\sigma(p)},j_{\tau(p)}},
\end{eqnarray*}
where $\sigma$ and $\tau$ run over the
$(k,N-k)$ shuffles.

To calculate $\det(\delta_{m_p,j_{\tau(q)}})$ we write
\begin{eqnarray*}
\det(\delta_{m_p,j_{\tau(q)}}) &=&
\sum_\lambda (-1)^\lambda \delta_{m_{\lambda(1)},j_{\tau(1)}}
\dots \delta_{m_{\lambda(k)},j_{\tau(k)}} ,
\end{eqnarray*}
where $\lambda$ runs over the permutations of $\{1,\dots,k\}$
and we obtain
\begin{eqnarray*}
\sum_{m_1,\dots,m_k}
\bara_{m_{1}}\dots \bara_{m_{k}}
\det(\delta_{m_p,j_{\tau(q)}}) &=&
k!
\bara_{j_{\tau(1)}}\dots \bara_{j_{\tau(k)}}.
\end{eqnarray*}
Hence
\begin{eqnarray*}
N^0_{KL} &=&
\sum_{k=0}^N (-1)^{k(k+1)/2}
\sum_{\sigma\tau} (-1)^{\sigma+\tau}
\alpha_{i_{\sigma(1)}}\dots \alpha_{i_{\sigma(k)}}
\\&&
\bara_{j_{\tau(1)}}\dots \bara_{j_{\tau(k)}}
\det(\delta_{i_{\sigma(p)},j_{\tau(q)}}).
\end{eqnarray*}

Therefore, our final result is
\begin{eqnarray}
N^0_{KL} &=&
\sum_{k=0}^N (-1)^{k(k-1)/2}
\sum_{\sigma\tau} (-1)^{\sigma+\tau}
\prod_{p=1}^k \bara_{j_{\tau(p)}}
\prod_{p=1}^k \alpha_{i_{\sigma(p)}}
\prod_{p=k+1}^N \delta_{i_{\sigma(p)},j_{\tau(p)}},
\nonumber\\
&=&
\sum_{k=0}^N
\sum_{\sigma\tau} (-1)^{\sigma+\tau}
\prod_{p=1}^k (\bara_{j_{\tau(p)}} \alpha_{i_{\sigma(p)}})
\prod_{p=k+1}^N \delta_{i_{\sigma(p)},j_{\tau(p)}},
\label{finalZKL}
\end{eqnarray}
where we recall that $\sigma$ and $\tau$
run over the $(k,N-k)$ shuffles.

We calculated $N^0_{KL}$ for a system where all the states
have the same number of electrons, but the same methods
can be used when $|K\rangle$ and $|L\rangle$ have
a different number of electrons. 
 
\subsection{Alternative formula for $N^0_{KL}$}
Now we are going to derive the alternative
formula (\ref{alternativex}) for $N^0_{KL}$.
This result can be obtained directly from 
equation(\ref{finalZKL}) but we shall provide
an independent proof.

We first rewrite the expression (\ref{an1ank})
for $N^0_{KL}$ as
\begin{eqnarray*}
N^0_{KL} &=& \sum_{k=0}^N \frac{(-1)^k}{(k!)^2}
\langle 0|b_{i_1}\dots b_{i_N} B_+^k B^k b^\dagger_{j_N}\dots b^\dagger_{j_1}|0\rangle,
\end{eqnarray*}
where $B=\sum_n \bara_n b_n$ and $B_+=\sum_n b^\dagger_n \alpha_n$.
It is easy to prove recursively that
$[ b_{i_N}, B_+^k]=k \alpha_{i_N} B_+^{k-1}$ and
$[ B^k\, b^\dagger_{j_N}]=k B^{k-1} \bara_{j_N}$.
If we write
$|K^-\rangle=b^\dagger_{i_{N-1}}\dots b^\dagger_{i_1}$,
so that $\langle K|=\langle K^-|b_{i_N}$,
we obtain the recursion
\begin{eqnarray*}
\langle K|B_+^k B^k|L\rangle &=&
\langle K^-| B_+^k b_{i_N}B^k|L\rangle + k \langle K^-| \alpha_{i_N} B_+^{k-1} B^k|L\rangle.
\end{eqnarray*}
If we use now $|L^-\rangle=b^\dagger_{j_{N-1}}\dots b^\dagger_{j_1}$, 
we obtain the following recursive equation between the matrix elements
of $B_+^k B^k$ for $N$ particles and $N-1$ particles:
\begin{eqnarray}
\langle K|B_+^k B^k|L\rangle &=&
\langle K^-|B_+^k B^k|L^-\rangle \delta_{j_N,i_N}
\nonumber\\&&
-k^2 \langle K^-|B_+^{k-1} B^{k-1}|L^-\rangle \bara_{j_N}\alpha_{i_N}
\nonumber\\&&
-k(-1)^{N-1} \langle K^-|B_+^{k-1} b^\dagger_{j_N} B^k|L^-\rangle \alpha_{i_N}
\nonumber\\&&
+k(-1)^{N-1} \langle K^-|B_+^{k} b_{i_N} B^{k-1}|L^-\rangle \bara_{j_N}
\nonumber\\&&
- \langle K^-|B_+^{k} b^\dagger_{j_N} b_{i_N} B^{k-1}|L^-\rangle.
\label{NLKrec}
\end{eqnarray}
Now we are going to show that the expression (\ref{alternativex}) satisfies
the same recursive equation.
We write $d=\sum_n\partial^2/\partial \alpha_n\partial \bara_n$,
$u=( \bara_{i_1} \alpha_{j_1}\dots \bara_{i_{N-1}} \alpha_{j_{N-1}})$
and $v=\bara_{i_N} \alpha_{j_N}$, so that
$N^0_{KL}=\ee^d(uv)$. Equations (\ref{d(uv)}) and (\ref{DeltaeD}) yield
\begin{eqnarray*}
\ee^d(uv) &=& \sum_{p=0}^\infty \frac{1}{p!} 
 d\i{1'}^{p} (\ee^d u)  d\i{2'}^{n} (\ee^d v).
\end{eqnarray*}
The sum is not infinite because
$\ee^d v=\delta_{j_N,i_N} + \bara_{j_N}\alpha_{i_N}$
so the sum stops at $p=2$.
Using
\begin{eqnarray*}
\Delta' d &=& \sum_n \frac{\partial}{\partial\alpha_n}\otimes \frac{\partial}{\partial\bara_n}
           -\frac{\partial}{\partial\bara_n}\otimes\frac{\partial}{\partial\alpha_n},
\\
\Delta' d^2 &=& \sum_{mn} 
-\frac{\partial^2}{\partial\alpha_m\partial\alpha_n}\otimes \frac{\partial^2}{\partial\bara_m\partial\bara_n}
-\frac{\partial^2}{\partial\bara_m\partial\bara_n}\otimes\frac{\partial^2}{\partial\alpha_m\partial\alpha_n}
\\&&
+\frac{\partial^2}{\partial\alpha_m\partial\bara_n}\otimes \frac{\partial^2}{\partial\bara_m\partial\alpha_n}
+\frac{\partial^2}{\partial\bara_m\partial\alpha_n}\otimes\frac{\partial^2}{\partial\alpha_m\partial\bara_n},
\end{eqnarray*}
we obtain the recursion
\begin{eqnarray}
\ee^d(uv) &=& (\ee^d u) \delta_{j_N,i_N} + (\ee^d u)\bara_{j_N}\alpha_{i_N}
+\frac{\partial (\ee^d u)}{\partial\alpha_{j_N}} \alpha_{i_N}
+\frac{\partial (\ee^d u)}{\partial\bara_{i_N}} \bara_{j_N}
\nonumber\\&&
+\frac{\partial^2 \ee^d u}{\partial\bara_{i_N}\partial\alpha_{j_N}}
\label{recurduv}
\end{eqnarray}

We use the derivatives
\begin{eqnarray*}
\frac{\partial B_+}{\partial \alpha_{j_N}} &=& - b^\dagger_{j_N},
\quad
\frac{\partial B_+^k}{\partial \alpha_{j_N}} = - k B_+^{k-1} b^\dagger_{j_N},
\end{eqnarray*}
to obtain
\begin{eqnarray*}
\frac{\partial \langle K^-| B_+^k B^k | L^-\rangle }{\partial \alpha_{j_N}} &=& 
-(-1)^{N-1} k \langle K^-| B_+^{k-1} b^\dagger_{j_N} B^k | L^-\rangle.
\end{eqnarray*}
Similarly
\begin{eqnarray*}
\frac{\partial \langle K^-| B_+^k B^k | L^-\rangle }{\partial \bara_{i_N}} &=& 
(-1)^{N-1} k \langle K^-| B_+^{k} b_{i_N} B^{k-1} | L^-\rangle,\\
\frac{\partial^2 \langle K^-| B_+^{k+1} B^{k+1} | L^-\rangle }{\partial \bara_{i_N}\partial \alpha_{j_N}} &=& 
(k+1)^2 \langle K^-| B_+^{k} b^\dagger_{j_N} b_{i_N} B^{k} | L^-\rangle.
\end{eqnarray*}
With these identities, it is easy to show that
$N^0_{K^-L^-}$ and $N^0_{KL}$ satisfy the same recursion
as $\ee^d(u)$ and $\ee^d(uv)$. 
When there is only one electron ($N=1$) it is easy to show
that $N^0_{KL}= \delta_{j_1,i_1} + \bara_{j_1}\alpha_{i_1}=\ee^d(\bara_{j_1}
           \alpha_{i_1})$.
Thus we have $N^0_{KL}=\ee^d(\bara_{j_1}\alpha_{i_1}\dots
   \bara_{j_N}\alpha_{i_N})$
for all $N$.

Notice that equation (\ref{recurduv}) enables us to 
calculate $N^0_{KL}$ for a closed shell, i.e. when
$N=M$. In that case there is only one state $|K\rangle$,
where all orbitals are filled, and we have
$i_k=j_k$ for all $k=1,\dots,N$. Since all the orbitals
are different, $e^d u$ does not contain $i_N=j_N$. Thus
the partial derivatives are zero and we obtain
$\ee^d(uv)=(\ee^d u)(1+\bara_{i_N}\alpha_{i_N})$.
For $N=1$ we have $N^0_{KL}=1+\bara_{i_1}\alpha_{i_1}$, thus
\begin{eqnarray}
N^0_{KL} &=& \prod_{k=1}^N (1+\bara_{i_k}\alpha_{i_k}).
\label{closedshell}
\end{eqnarray}

\end{document}